\documentclass[a4paper]{amsart} 
\usepackage{amsmath}
\usepackage{amssymb}
\usepackage{verbatim}
	\usepackage{amsthm}
	\usepackage{mathrsfs}
	\usepackage{graphicx}
	\usepackage{mathtools}
	\usepackage[colorlinks,pdfdisplaydoctitle,linkcolor=blue,citecolor=blue]{hyperref}
	\usepackage{caption}
	\usepackage{subcaption}
	\usepackage{url}
	\usepackage{epstopdf}
	\usepackage{pgf,tikz}
	\usepackage{bbm}
	\usepackage{extarrows}
	\usetikzlibrary{arrows}
	\usepackage{algorithm}
	\usepackage{algorithmicx}
	\usepackage{tikz}
	\usepackage{pgfplots}
	\usetikzlibrary{intersections}
	\usetikzlibrary{shapes.misc, pgfplots.groupplots,pgfplots.fillbetween}
	\usepackage[natbib=true,bibstyle=numeric,backend=bibtex,giveninits=true,maxbibnames=9, maxcitenames=5]{biblatex}
	\DeclareNameAlias{author}{family-given}
	
	\usepackage{ dsfont }
	\usepackage{listings}
	\usepackage{bbm}
	\usepackage{color}
	\usepackage{ dsfont }
	\usepackage{enumerate}
	\usepackage{todonotes}

	\renewcommand{\Delta}{\triangle}

	\definecolor{darkblue}{rgb}{0,0,0.7}

	\definecolor{darkgreen}{rgb}{0.01,0.75,0.24}

	\def \Ee[#1]{\mathcal{E}^{\text{{#1}}}}

	\def\pa[#1,#2]{\frac{\partial {#1}}{\partial {#2}} }
	
	\def\idom[#1,#2,#3]{\int_{#1}\hspace{1pt} {#2} \hspace{1pt} \text{d}{#3}}
	\def\res[#1,#2]{\left.{#1}\right|_{#2}}

	\def\var[#1,#2]{\langle \delta \mathcal{E}^{\text{{#1}}}({#2}),v\rangle}
	\def\vars[#1,#2,#3]{\langle \delta^2\mathcal{E}^{\text{{#1}}}({#2})v,{#3}\rangle}
	\def\vard[#1,#2,#3,#4]{\langle \delta\mathcal{E}^{\text{{#1}}}({#2})-\delta\mathcal{E}^{\text{{#3}}}({#4}),v\rangle}

	\def\P{\mathbb{P}}
	
	\def\E{\mathbb{E}}

	\newcommand{\bm}{\mathbf{m}}

	\newcommand{\be}{\begin{equation}}
		\newcommand{\en}{\end{equation}}
	\newcommand{\ben}{\begin{equation*}}
		\newcommand{\enn}{\end{equation*}}
	\newcommand{\bea}{\begin{aligned}}
		\newcommand{\ena}{\end{aligned}}
	
	\def\ba#1\ena{\begin{align}#1\end{align}}
	\def\ban#1\enan{\begin{align*}#1\end{align*}}

	\theoremstyle{plain}
	
	\newtheorem{theorem}{Theorem}[section]

	\numberwithin{equation}{section}

	\DeclareMathAlphabet{\bi}{OML}{cmm}{b}{it}

	\newcommand{\defeq}{\mathrel{\mathop:}=}
	\newcommand{\eqdef}{=\mathrel{\mathop:}}
	\newcommand{\real}{\mathbb{R}}
	\newcommand{\complex}{\mathbb{C}}
	
	\newcommand{\nanu}{\mathbb{N}}
	\renewcommand{\P}{\mathbb{P}}
	\newcommand{\stable}[2]{\mathcal{S}_{#1}( #2 )}

	\newcommand{\rme}{\mathrm{e}}
	\newcommand{\rmi}{\mathrm{i}}
	\newcommand{\rmd}{\mathrm{d}}
	\newcommand{\fl}{}
	
	\newcommand{\eref}[1]{(\ref{#1})}
	\newcommand{\remainder}{\mathcal{R}}
	\newcommand{\eqalign}{}
	
	\usepackage{xcolor}
	\usepackage{graphicx}
	\usepackage{bm}
	\usepackage{tikz}
	\usepackage{pgfplots}
	\usepackage{subcaption}
	\pgfplotsset{compat=newest}
	\usepackage{placeins}

\begin{document}
		
		\title[Bayesian inversion with $\alpha$-stable priors]{Bayesian inversion with $\alpha$-stable priors}

		\author[J. Suuronen] {Jarkko Suuronen}
		\address{School of Engineering Science, Lappeenranta-Lahti University of Technology, FI-53850, Finland}
		\email{jarkko.suuronen@lut.fi}
		
		\author[T. Soto] {Tomas Soto}
		\address{School of Engineering Science, Lappeenranta-Lahti University of Technology, FI-53850, Finland}
		\email{tomas.soto@lut.fi}
		
		\author[N. K. Chada] {Neil K. Chada}
		\address{Department of Actuarial Mathematics and Statistics, Heriot Watt University, Edinburgh, EH14 4AS, UK}
		\email{neilchada123@gmail.com}
		
		\author[L. Roininen] {Lassi Roininen}
		\address{School of Engineering Science, Lappeenranta-Lahti University of Technology, FI-53850, Finland}
		\email{lassi.roininen@lut.fi}

		
		\begin{abstract}
			
			We propose using Lévy $\alpha$-stable distributions to construct priors for Bayesian inverse problems.
            The construction is based on Markov fields with stable-distributed increments. 
            Special cases include the Cauchy and Gaussian distributions, with stability indices $\alpha=1$, and $\alpha=2$, respectively. 
            Our target is to show that these priors provide a rich class of priors for modeling rough features.
            %
            %
            The main technical issue is that the $\alpha$-stable  probability density functions lack closed-form expressions, and this limits their applicability. 
            For practical purposes, we need to approximate probability density functions through numerical integration or series expansions.
            For Bayesian inversion, the currently available approximation methods are either too time-consuming or do not function within the range of stability and radius arguments. 
            To address the issue, we propose a new hybrid approximation method for symmetric univariate and bivariate $\alpha$-stable distributions that is both fast to evaluate and accurate enough from a practical viewpoint.
            In the numerical implementation of $\alpha$-stable random field priors, we use the constructed approximation method. 
            We show how the constructed priors can be used to solve specific Bayesian inverse problems, such as the deconvolution problem and the inversion of a function governed by an elliptic partial differential equation.
            We also demonstrate hierarchical $\alpha$-stable  priors in the one-dimensional deconvolution problem. 
            For all numerical examples, we use maximum a posteriori estimation.
            To that end, we exploit the limited-memory BFGS and its bounded variant for the estimator. 
			%

		\end{abstract}

		\maketitle

    \section{Introduction}\label{sec:intro}
    
    Inverse problems is the mathematical theory and practical interpretation of noise-perturbed indirect observations. 
    Bayesian statistical inversion is the effort to formulate real-world inverse problems as Bayesian statistical estimation problems \cite{AT87,KS04}. Bayesian inverse problems can be found in medical and subsurface imaging, industrial applications, and near-space remote sensing. The objective, for example in industrial tomography, is to detect different materials, which may have isotropic, anisotropic, or inhomogeneous features. This means that we typically aim to reconstruct a hidden substance from indirect noise-perturbed measurements. Inhomogeneities include, for example, material interfaces and rough features, and these are the main topics of this paper.
    
    Inverse problems are often formulated using a noise-perturbed measurement equation
    \begin{equation}
        \label{eq:ip}
        \mathbf{y}=\mathcal{G}(u) + \bm{\eta}, \quad \bm{\eta} \sim \mathcal{N}(\mathbf{0},\bm{C}),
    \end{equation}
    where $\mathbf{y} \in \real^M$ are noisy  finite-dimensional measurements, $\mathcal{G}$ is a linear or non-linear mapping from some function space to $\real^M$, $u: D \rightarrow \real$ is the unknown with $D \subseteq \real^d,\, d = 1,2,3$,  and $\bm{\eta}$ is noise, which we assume to be Gaussian.  
    Our aim is to estimate $u$ given one realization of $\mathbf{y}$.
    
    Inverse problem methods can be roughly divided into deterministic and statistical methods.
    We model $\mathbf y, u, \bm{\eta}$ as random objects in a statistical framework \cite{KS04}. 
    For practical computations, we discretize the unknown $u$, and denote it by $\mathbf{u}$.
    Then, within the Bayesian inverse problem framework, the solution can be represented through probability distributions via Bayes' theorem, that is, the posterior distribution	\begin{equation}
        \label{eq:post}
            \pi(\mathbf{u}\mid \mathbf{y}) = \frac{\ \pi( \mathbf{y}\mid \mathbf{u}) \pi(\mathbf{u})}{\pi( \mathbf{y}) } \nonumber 
            \propto \pi( \mathbf{y}\mid \mathbf{u}) \pi(\mathbf{u}), 
    \end{equation}
    where $\pi( \mathbf{y}\mid \mathbf{u})$ is the likelihood, and  $\pi( \mathbf{u})$ is the prior distribution of the unknown. We omit the normalization constant $\pi( \mathbf{y})$, and instead use the unnormalized posterior distribution from now on.
    
    The choice of the prior $\pi(\mathbf u)$ is practically the only tuneable object in inversion.
    The traditional choices in inverse problems are Gaussian and total variation priors for smoothing and edge-preserving inversion, respectively \cite{KS04,RHL14,RW06,LS04}.  
    In this paper, we build upon the research line starting from the observation that total variation priors do not provide invariant estimators under mesh refinement \cite{LS04}. 
    Besov priors on a wavelet basis were proposed as a solution to this problem  \cite{LSS09}.
    Here, we extend the study from Cauchy \cite{SCLR21} priors to $\alpha$-stable priors, of which the Cauchy priors are special cases with $\alpha=1$, and Gaussian priors are similarly special cases with $\alpha=2$.
    In order to leverage $\alpha$-stable laws for Bayesian inverse problems, we need approximations of  $\alpha$-stable probability densities evaluated very fast with reasonable precision \cite{N97,JPN13}. 
    Our particular interest is to implement and use discretized $\alpha$-stable random fields in Bayesian continuous-parameter estimation.

    We note that traditionally, stable distributions have been employed in financial applications like the modeling of asset time series \cite{BSS21}. They have also been used in biomedical engineering \cite{ABTTK15}, remote sensing \cite{MA11}, network traffic statistical analysis \cite{XGY04}, and digital signal processing \cite{NS95}, to name a few. 
    We extend the application to inverse problems here.

    \subsection{Contributions}
    
    Our main contribution is to implement computationally feasible numerical approximations of symmetric $\alpha$-stable priors for Bayesian inverse problems, which requires evaluating the univariate or multivariate probability density functions of $\alpha$-stable random variables. 
    The symmetric $\alpha$-stable probability density functions do not have elementary function expressions, except for the two special cases of Gaussian and Cauchy distributions. 
    Thus, the evaluation of the probability densities requires the incorporation of an appropriate approximation method.

    Unfortunately, none of the existing $\alpha$-stable density function approximation methods is alone optimal for our needs. 
    Being based on either series expansions, integral expressions or Fourier transforms \cite{  JPN13,TS94,CR18,N97,AO18,MR06,CAMM18,TA08}, they are either computationally too heavy to evaluate within Bayesian inversion,  are  not applicable for arbitrary values of $r$ and stability indices $\alpha$, or do not provide a consistent seamless approximation in the sense of the partial derivatives. 
    
    For this reason, we introduce a fast hybrid method to approximate the $\alpha$-stable laws that leverage both bicubic spline interpolations at precomputed probability density grids and asymptotic series approximations. 
    The novelty of the hybrid method is multi-part. 
    First, the presented method is orders of magnitude faster to evaluate than the prevalent methods that are based on numerical integration. 
    The method approximates the log densities of the $\alpha$-stable distributions within 100-400 nanoseconds on a typical Intel Xeon-based workstation. 
    For the second, the method enables the evaluation of the partial derivatives of the log densities with respect to both $\alpha$ and $r$ with similar performance, which is crucial for the estimators in the Bayesian inference. 
    Finally, the method provides a consistent and discontinuity-free approximation of the log densities on a wide range of stability indices $\alpha$ and for any argument $r$ as opposed to the series expansion methods.  
    
    We  establish error bounds for the method through comprehensive computer-assisted analysis of the partial derivatives of the $\alpha$-stable probability density functions \cite{MT04} and argue that the presented approach is accurate enough for Bayesian inversion. 
    Furthermore, we demonstrate various $\alpha$-stable priors on a range of Bayesian inverse problems. To our knowledge, general $\alpha$-stable priors have not been approximated and implemented numerically for Bayesian inverse problems previously. 
    The numerical examples include deconvolution problems in one-dimensional and two-dimensional grids. Additionally, we illustrate nonlinear Bayesian inversion governed by an elliptic partial differential equation through $\alpha$-stable priors. In the numerical examples, we resort to maximum a posteriori (MAP) estimators: 
    \begin{equation}
        \label{mappi}
        \mathbf{u}_{\rm{MAP}} :=  \arg \max_{\mathbf{u}}	\pi(\mathbf{u}\mid \mathbf{y}).
    \end{equation}
    We showcase the applicability of the presented methodology in a hierarchical $\alpha$-stable prior, of which the stability index $\alpha$ or the scale $\sigma$ are processes of their own. 
    The presented hybrid method proved to be very useful in such a hierarchical scenario because of the variable stability indices.  
    In essence, we extend previous works on parametric deep Gaussian processes \cite{ZES} to simple two-layer $\alpha$-stable processes.   
    The  $\alpha$-stable priors have the ability to model both discontinuities and smoothness simultaneously in a function of interest, which is a desired property in Bayesian continuous-parameter estimation. 
    The results are promising for further development of $\alpha$-stable random field priors. For instance,  the hierarchical $\alpha$-stable priors may be very effective tools to model processes that have simultaneously properties of a stationary Gaussian process and a Cauchy random walk.
    
    
    
    \subsection{Outline}
    
    This paper is organized as follows: 
    We provide the necessary background material for the paper as an introduction to $\alpha$-stable priors in Section  \ref{sec:models}.
    This will lead onto Section \ref{sec:error_bound}, where we briefly review the existing methods and our hybrid method for approximating $\alpha$-stable probability density functions, and provide error bounds related to our method. 
    Numerical experiments with the $\alpha$-stable priors are provided in Section \ref{sec:num}, where we test our priors on  the example problems. A summary of our findings and future work is provided in Section \ref{sec:conc}.

    \section{Models}\label{sec:models}
    
    In this section, we review and discuss the necessary prior forms based on $\alpha$-stable distributions. We also present some basic properties and then present the multivariate setting and how they can be defined. 
    
    \subsection{Stable distributions}\label{ssec:distributions}
    A random variable $W$ corresponding to a symmetric \emph{stable} distribution, also known as \emph{$\alpha$-stable} and \emph{Lévy $\alpha$-stable} distribution, can be characterized in terms of a \emph{stability index} $\alpha \in (0,2]$ (sometimes also called the \emph{tail index} or the \emph{characteristic exponent}), and a scale parameter $\sigma > 0$, in the sense that its characteristic function is given by
    \begin{equation}\label{eq:stable-cf}
        \E\left[ \exp(\rmi \theta W) \right] = \exp\left(-(\sigma |\theta|)^\alpha\right), \quad \theta \in \real,
    \end{equation}
    in which case we write
    \[
    W \sim \stable{\alpha}{\sigma}.
    \]
    The parameter $\alpha$ is called the stability index because if $W_1$ and $W_2$ are two independent copies of $W$ and $A$, $B > 0$, then
    \begin{equation}\label{eq:stable-distribution}
        A W_1 + B W_2 \stackrel{d}{=} C W
    \end{equation}
    with
    \[
    C^\alpha = A^\alpha + B^\alpha.
    \]
Hence, the symmetric $\alpha$-stable distributions are a family of continuous probability distributions that are infinitely divisible and closed under convolution. The monograph \cite{TS94} is the standard reference for stable distributions, including the wide class of non-symmetric stable distributions, which we do not consider here. 

It is clear from \eref{eq:stable-cf} that for $\alpha = 2$, $W$ is normally distributed (with zero mean and variance $2\sigma^2$). Likewise, for $\alpha = 1$, $W$ has a Cauchy distribution (with zero median and scale parameter being $\sigma$). Besides these two special cases of $\alpha$-stable laws, there are no known closed-form  expressions based on elementary functions for the density functions of symmetric stable distributions (the other special cases where closed-form expressions are known to consist of non-symmetric distributions, such as the Lévy distribution).

It holds that $\E[|W|^\alpha] = \infty$ for $\alpha < 2$, implying that an $\alpha$-stable distribution has infinite variance and that its mean is not well-defined  for $\alpha \leq 1$. It does, however, hold that $\E[|W|^\lambda] < \infty$ for all $\lambda \in (0,\alpha)$.  

Multivariate stable distributions can be defined in a similar but more complicated manner, with a \emph{spectral measure} $\Lambda$ in place of the scale parameter $\sigma$; see \cite[Chapter 2]{TS94}. %
For our purposes, it suffices to recall the definition of \emph{spherically contoured stable distributions}: a random vector $\bi{W} \eqdef (W_1,W_2,\cdots,W_d)$ on $\real^d$ is said to have a spherically contoured stable distribution if its characteristic function is given by
\[
\E\left[ \exp\left(\rmi \sum_{j=1}^d \theta_j W_j \right) \right] = \exp\left(-(\sigma |\theta|)^\alpha\right), \quad \theta \in \real^d,
\]
where $\alpha \in (0,2]$ is again a stability index, $\sigma > 0$ is a scale parameter and $|\cdot|$ stands for the standard $\ell^2$-based Euclidean norm on $\real^d$. We refer to \cite{JPN13} for a treatment of such distributions. 



\subsection{$\alpha$-stable priors}

A stochastic process $(W_t)_{t \geq 0}$ is a symmetric $\alpha$-stable process if $\sum_{j=1}^n a_j W_{t_j}$ is a symmetric $\alpha$-stable random variable for all finite $\{t_1,\cdots,t_n\} \subset [0,\infty)$ and $\{a_1,\cdots,a_n\} \subset \real$.
We refer to \cite[Chapter 3]{TS94} for the existence and construction of a wide variety of such processes. 
If the previous definition is satisfied for the multivariate case $\{t_1,\cdots,t_n\} \subset  \real^K$, the process is called an $\alpha$-stable field.

In particular, we aim to apply discretized priors corresponding to a \emph{L\'evy $\alpha$-stable motion}, which we take to mean an $\alpha$-stable process $(W_t)_{t \geq 0}$ with some given initial distribution $W_0 \sim \mu$ and independent increments that satisfies
\[
W_t - W_s \sim \stable{\alpha}{|t-s|^{1/\alpha}}, \quad \textrm{for all~}  s,\,t \in [0,\infty), \, t \neq s.
\]
For $\alpha < 2$, the L\'evy $\alpha$-stable motion generally does not have continuous sample paths. However, according to \cite[Theorem 11.1]{KS99}, there exists a version of this process with c\'adl\'ag paths satisfying
\begin{equation}\label{eq:levy-path-continuity}
\P( W_t = W_{t-}) = 1, \quad \textrm{for all~}  t > 0.
\end{equation}
We more generally refer to \cite{KS99} for an overview of the analytical properties of L\'evy $\alpha$-stable motions and related processes, including a description of their infinitesimal generators.

A L\'evy $\alpha$-stable motion with initial distribution $\mu$ can be discretized as follows. For $\Delta \in (0,1)$, define the Markov chain $(u^\Delta_k)_{k \geq 0}$ by $u^\Delta_0 \sim \mu$, and $u^\Delta_{k+1} - u^\Delta_k \sim \stable{\alpha}{\Delta^{1/\alpha}}$ independently for all $k \geq 0$. Then, by writing $(W^\Delta_t)_{t \geq 0}$ for the appropriately-scaled, piecewise constant c\'adl\'ag process stemming from the Markov chain $(u^\Delta_k)_{k \geq 0}$, i.e.
\[
W^\Delta_t \defeq u^\Delta_{\lfloor t/\Delta\rfloor}.
\]
Using the basic properties of stable distributions along with \eref{eq:levy-path-continuity}, it is easy to verify that $\lim_{\Delta\to 0^+} W^\Delta = W$ in the sense of finite-dimensional distributions, i.e.
\[
\lim_{\Delta\to 0^+} \E\bigl[h\bigl(W^\Delta_{t_1},\cdots,W^\Delta_{t_n}\bigr)\bigr] = \E\bigl[h\bigl(W_{t_1},\cdots,W_{t_n}\bigr)\bigr],
\]
for all finite $\{t_1,\cdots,t_n\} \subset [0,\infty)$ and bounded and continuous functions $h\colon \real^n\to\real$.


An alternative way to construct such a discretization, localized to a finite interval, is to partition the interval by $N$ equispaced points and define the unnormalized density function of $\mathbf{u} \defeq (u_i)_{i=1}^{N}$ on these points as
\label{process}
\begin{align}
    & \pi(\mathbf{u})  \propto \mu(u_1) \prod_{i=2}^N f( u_i - u_{i-1}; \alpha, \sigma), 
\end{align}
where $\mu$ is the initial distribution of the above-mentioned process $(W_t)_{t\geq 0}$  and $f(\, \cdot \,;\alpha, \sigma)$ stands for the stable density function with stability index $\alpha$ and appropriately-chosen scale parameter $\sigma$.

The only two-dimensional $\alpha$-stable field we consider in this paper is a simple generalization of the quasi-isotropic Cauchy first-order difference prior \cite{SCLR21}, defined analogously to \eref{process}. That is, the probability density function of an $\alpha$-stable random field $\mathbf{u}$ discretized through finite differences on a two-dimensional rectangular domain $\Omega \subset \real^2$ is proportional to
\begin{equation}
\label{bivariate}
\pi(\mathbf{u}) \propto \pi_{\partial \Omega}(\mathbf{u}_{\partial \Omega}) \prod_{i,j \notin\partial \Omega } f_B(u_{i,j}-u_{i,j-1},u_{i,i}-u_{i-1,j}; \alpha; \sigma),
\end{equation}
where $\partial \Omega$ denotes the set of the left and bottom indices on the grid, and $f_B(\cdot,\cdot; \alpha, \sigma)$ the symmetric bivariate $\alpha$-stable probability density function. The probability density function $ \pi_{\partial \Omega}$ is applied on the grid points at the left and bottom boundary of the grid to make the resulting distribution of $\mathbf{u}$ proper.

\subsection{Hierarchical $\alpha$-stable priors}

Hierarchical priors are dominantly used within Gaussian priors \cite{DL13, ZES,ARSH20}.
With these priors, we can model discontinuities and other features with varying scale or smoothness at the target function. 
Unfortunately, the computational complexity of the canonical  Gaussian priors is cubic with respect to the number of  training points, unless a special formulation of the process is employed, like a stochastic partial differential equation \cite{LRL11}. 
The hierarchical priors might require several layers on top of each other to perform well, while having too many layers may not offer any additional expression capability \cite{DGS18} but rather overfit in the data.

We intend to build and demonstrate simple, two-layer Markovian hierarchical $\alpha$-stable priors that could be useful without the computational or implementation complexity of hierarchical Gaussian processes.
We model the scale or the stability of a discretized $\alpha$-stable process as another $\alpha$-stable process. 
This is possible due to the first-order difference prior's simple Markovian construction, which effectively allows expressing the normalization constant of the joint distribution of the discretized process and its parameters in a closed form.
A hierarchical $\alpha$-stable difference process $\mathbf{u}$ with scale $\sigma = G(c)$ and stability $\alpha = H(s)$ based on other discretized $\alpha$-stable difference processes could be constructed as follows:
\begin{align}\label{hiera}
    & \pi(\mathbf{u},\mathbf{c}, \mathbf{s}\mid \mathbf{y}) \propto  \pi( \mathbf{y} \mid \mathbf{u}) \pi(\mathbf{u}\mid \mathbf{c},\mathbf{s}) \pi(\mathbf{c}) \pi(\mathbf{s}) =  \\ 
    &\pi(\mathbf{y}\mid \mathbf{u}) f\left( u_1; H(s_1), G(c_1)  \right)  f( c_1; \alpha_c, \sigma_c  ) f( s_1; \alpha_s, \sigma_s  ) \times \\ &  \prod_{i=2}^N f\left( u_i - u_{i-1}; H(s_i), G(c_i)   \right)  f( c_i-c_{i-1};\alpha_c, \sigma_c ) f( s_i-s_{i-1}; \alpha_s, \sigma_s ), 
\end{align}
where $H$ and $G$ are nonlinear functions with  $\textrm{Range}(G) \subseteq \real^+$, and $ \textrm{Range}(H) \subseteq (0,2]$.   $f(\cdot;\alpha,\sigma)$ denotes the probability density function of a symmetric univariate $\alpha$-stable random variable with stability $\alpha$, scale $\sigma$. 
The conditional distribution $\pi(\mathbf{u} \mid \mathbf{c},\mathbf{s})$ integrates to a constant, regardless of $\mathbf{c}$ and $\mathbf{s}$, as do the priors $\pi(\mathbf{c}) $ and $ \pi(\mathbf{s})$, because $\sigma_s,\sigma_c,\alpha_s$ and $\alpha_c$ are fixed. The overall joint prior distribution is thus proper.

The convergence properties to the continuous limit of the hierarchical $\alpha$-stable processes in the continuous-time limit are unknown -- a sum of two Lévy $\alpha$-stable random variables with different stability indices does not obey an $\alpha$-stable distribution. However, continuous-time processes with a local stability index varying with the state of the process, commonly called \emph{stable-like processes}, are well-studied in the literature; see e.g. Chapter 7 in \cite{KO11} and  Theorem 5.2 in \cite{KU17}.

Further studies are thus needed regarding the matter, but as we demonstrate in the numerical experiments, the hierarchical $\alpha$-stable process constructions are promising. 
Unfortunately, the presented hierarchical priors  cannot be applied to the $\alpha$-stable difference  priors when the spatial dimension is greater than one. 
That is because the normalization constants of the priors are intractable due to their construction upon the distributions of increments between nearest neighbors. However, a Matérn-like stochastic partial differential equation prior could be optionally employed instead of the difference priors \cite{SCLR21}, which would effectively allow incorporating deep $\alpha$-stable processes thanks to the tractable normalization constants. 


\section{Approximation of $\alpha$-stable probability density functions}
\label{sec:error_bound}
Before presenting our hybrid method of approximating symmetric spherically-contoured $\alpha$-stable density functions, we perform a brief literature overview of the existing approximation methods as a motivation for our contributions.  
Later, we provide various relative error bounds for the probability density approximations given by our method. For simplicity, we denote with $r$ both the argument of the univariate probability density functions and the Euclidean distance of the arguments of the multivariate $\alpha$-stable probability density functions to the origin, that is, $r:= \Vert\mathbf{r}\Vert$.

Unless otherwise specified, the approximations are applied for $\sigma=1$. Recall that for general symmetric $\alpha$-stable laws, the probability density functions for the other scale parameters are given by $f(r;\alpha,\sigma) = \frac{1}{\sigma^d}f(\frac{r}{\sigma};\alpha,1) $, where $d$ stands for the dimensionality of the distribution \cite{JPN13}.

\subsection{Prevalent approximation methods for symmetric univariate and bivariate $\alpha$-stable laws}
%

A canonical method to approximate the $\alpha$-stable probability density functions is to evaluate the inverse Fourier transform of the characteristic function. For the symmetric univariate $\alpha$-stable distributions given by \eref{eq:stable-cf} with $\sigma = 1$, the probability density function can be expressed  as \cite{N99}
\begin{equation}\label{direct}
f(r;\alpha) = \frac{1}{\pi} \int_0^{\infty}  \cos(rt) \exp(-t^{\alpha}) \rmd t.
\end{equation}
In theory,  numerical integration allows evaluating the density of any $\alpha$-stable distribution at an arbitrary point $r$ with specified precision. The integral may be impractical to evaluate for large $r$ and small $\alpha$ due to the severe oscillations \cite{N99}, so oscillatory integral techniques have been proposed to address the issue \cite{AO18}.  Furthermore, the discrete Fourier transform method \cite{SR09} exploits the low computational complexity of the fast Fourier transform, but requires interpolation to evaluate the density outside the grid \cite{MDC99, MR06}.

Additionally, there is an alternative integral representation formula \citep{N97}, that we call Nolan's method for short, for the univariate $\alpha$-stable density function when $\alpha \neq 1$. For simplicity, if we consider only symmetric $\alpha$-stable distributions and  $\alpha \neq 1$, the probability density function for an $\alpha$-stable random variable with $\mu=0$ and $\sigma=1$ given by the method is \cite{N97}
\begin{equation}\label{nolanintegral}
f(r;\alpha) = \frac{\alpha |r|^{\frac{1}{\alpha-1}}}{\pi|\alpha-1|} \int_0^{\pi/2} Q(t,\alpha) \exp\left(-|r|^{\frac{\alpha}{\alpha-1} } Q(t,\alpha)\right) \rmd t,
\end{equation} 
where $Q(t,\alpha) =  \left( \frac{\cos(t)}{\sin(\alpha t )} \right)^{\frac{\alpha}{\alpha-1}} \frac{\cos(\alpha t -t)}{\cos (t)}$. 
In contrast to the inverse Fourier method,  the integrand in Nolan's method is compactly supported and non-oscillatory. 
Unfortunately, when $|\alpha-1| < 0.02$, the integrand becomes extremely peaky and narrow, and then the method is impractical unless arbitrary precision arithmetic is used for the integration \cite{N97}. Even otherwise  Nolan's method   can be tricky to evaluate since the domain of integration should be evaluated in parts near the peak of the integrand.
The peak is located at $t_p$, which satisfies the equation $Q(t_p,\alpha) |r|^{\frac{\alpha}{\alpha-1}} = 1$ \cite{N97}. 
Thus, the univariate symmetric $\alpha$-stable laws can be accurately evaluated with either Nolan's method or the inverse Fourier transform method depending on the values of $\alpha$ and $r$. The approximation methods based on the Fast Fourier transform \cite{SR09, B05, MR06, MDC99} are also worth mentioning. They are simple to implement and relatively fast to evaluate but must be used in conjunction with interpolation to approximate the density at a point that is not part of the FFT grid. It has  been reported that the  FFT-based approximation is accurate only for large $\alpha$  \cite{BHW05,MR06}.

The  approximations based on the integral representations of $\alpha$-stable laws are complemented by series expansions.     A well-known series expansion for the univariate $\alpha$-stable density is of form \cite{HB52}
\begin{equation}\label{tail}
f(r;\alpha) \sim -\frac{1}{\pi} \sum_{k=1}^{\infty} \frac{(-1)^k \Gamma(k\alpha + 1) \sin(\frac{k\pi\alpha}{2})}{k!} r^{-k\alpha-1},
\end{equation}
which is an asymptotic expansion for $\alpha \in (1,2)$ for $r \to \infty$, and converges pointwise to the true density for $\alpha \in (0,1)$.  
There is a similar series expansion, also outlined in \cite{HB52}, which is an asymptotic series for $\alpha \in (0,1)$ and a converging series for $\alpha \in (1,2)$ at $ r\to 0^+$. 
Furthermore, there are methods that provide a converging series approximation for the symmetric univariate probability densities for  $\alpha \in (0,2)$  by combining two separate power series expansions \cite{CAMM18}, or approximate the inverse Fourier transform of the characteristic function by domain partitioning and implementing different series expansions within them \cite{CR18}.

The methodology for approximating spherically contoured multivariate $\alpha$-stable distributions is similar to the univariate one. There are several integral expressions for their probability density functions (see e.g.~\cite{JPN13}), such as
\begin{align}
\label{multivar}
    f_M(r;\alpha_M) & =  \frac{2^{1-d/2}}{\Gamma(d/2)} \int_0^{\infty} (rs)^{\frac{d}{2}} J_{d/2-1} (rs) \exp{\left ( - s^{\alpha}\right)} \rmd s
\end{align}
where $J_\nu$ is the Bessel function of the first kind. 
Analogously to the univariate case, multivariate spherically contoured $\alpha$-stable laws have an absolutely converging  series expansion  for $r > 0$ and $\alpha \in (0,1)$, which is an asymptotic expansion for $\alpha \in (1,2)$ and $r \to \infty$ \cite{JPN13}:
\begin{equation}
\label{multiseriesinfty}
\eqalign{
    f_M(r;\alpha) \sim \frac{-1}{2\pi^{d/2+1} } \sum_{k=1}^{\infty} \frac{(-1)^{k}\Gamma(\frac{k\alpha+2}{2})\Gamma(\frac{k\alpha + d}{2})\sin( \frac{k\alpha \pi}{2})}{ k!}\Bigl( \frac{r}{2}\Bigr)^{-(k\alpha+1)}.}
\end{equation}
Similarly, there is an absolutely converging series expansion for $r > 0$ and $\alpha \in (1,2)$, which is an asymptotic expansion for $\alpha \in (0,1)$ and $r\to 0^+$ \cite{JPN13}.

None of the existing approximation methods as such is suitable to be used in Bayesian inversion. 
The presented approximation techniques based on numerical integration are not applicable for the tails of the $\alpha$-stable distributions because the relative accuracy of the integration cannot be ensured due to limited floating point precision. They and the advanced series expansions \cite{CR18} may be too time-consuming to perform within Bayesian continuous-parameter estimation since the probability density functions must be evaluated up to several hundreds of thousands of times even in modest-dimensional settings.  Some of the methods are also applicable only on fixed $\alpha$ \citet{MR06}, and not on a continuous range of them. Certain  approximation methods fail to approximate the log density of the $\alpha$-stable distributions with a close to uniform relative accuracy for all $r$ because they are based on combining two power series expansions \cite{CAMM18}, which may affect negatively the estimators. Finally, the typical asymptotic series expansions are fast to evaluate when the number of used terms is low, but not accurate enough or even applicable for all $r$, let alone $\alpha$.   

\subsection{Hybrid method for approximating $\alpha$-stable laws}

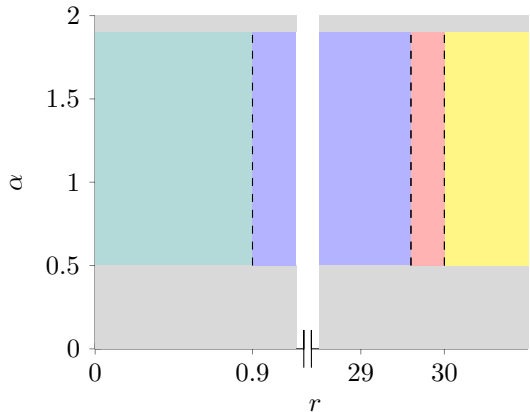
\begin{figure}[ht]
\centering
\begin{subfigure}[b]{0.75\textwidth}
    \centering     
    \begin{tikzpicture}[]
        \centering			
        \begin{groupplot}[
            group style={
                group size=2 by 1,
                yticklabels at=edge left,
                horizontal sep=3mm,
            },
            height=6cm,
            ymin=0, ymax=2,
            domain=0:35,
            ]
            
            \nextgroupplot[
            xmin=0,xmax=1.15,
            xtick={0,0.9},
            ytick = {\empty},
            extra y ticks={0,0.5,1,1.5,2},
            extra y tick labels={0,0.5,1,1.5,2},
            axis y line=left,
            axis x line=bottom,
            y axis line style=-,
            ylabel = {$\alpha$},
            x axis line style=-, 
            x=2.3cm,      
            ]
            \addplot [name path = lahi0,
            dashed,
            domain = 0:1.15]  coordinates {(0.9,0.5) (0.9,1.9)};
            
            \addplot[name path=reuna0, mark=none, draw=none] coordinates {(0,0.0) (0,0.5)};
            \addplot[name path=reuna1, mark=none, draw=none] coordinates {(1.15,0.0) (1.15,0.5)};
            
            \addplot[name path=reuna2, mark=none, draw=none] coordinates {(0,1.9) (0,2)};
            \addplot[name path=reuna3, mark=none, draw=none] coordinates {(1.15,1.9) (1.15,2)};
            
            
            \addplot[name path=nolla, mark=none, draw=none] coordinates {(0,0.5) (0,1.9)};
            
            \addplot [teal!30] fill between [of = lahi0 and nolla,    reverse=true];
            
            \addplot [gray!30] fill between [of = reuna0 and reuna1,    reverse=true];
            \addplot [gray!30] fill between [of = reuna2 and reuna3,    reverse=true];
            
            \addplot[name path=loppuno, mark=none, draw=none] coordinates {(1.15,0.5) (1.15,1.9)};
            
            \addplot [blue!30] fill between [of = lahi0 and loppuno,  reverse=true];
            
            \coordinate(dl)at(rel axis cs:1,0);
            
            \nextgroupplot[
            xmin=28.5,xmax=31,
            axis y line=none,
            xtick={29,30},
            xlabel = {\hspace{-2.9cm}$r$},    
            xlabel style={below right}, 
            axis x line=bottom,
            x axis line style=-, 
            x=1.1cm,      
            ]
            
            \coordinate(dr)at(rel axis cs:0,0);
            
            \addplot[name path=ensi,  mark=none, dashed] coordinates {(29.6,0.5) (29.6,1.9)}; 
            
            \addplot[name path=alkur,  mark=none, draw=none]  coordinates {(13,0.5) (13,1.9)};

            \addplot [blue!30] fill between [of = ensi and alkur,    reverse=true];
            
            \addplot[name path=reuna4, mark=none, draw=none] coordinates {(28.5,0.0) (28.5,0.5)};
            \addplot[name path=reuna5, mark=none, draw=none] coordinates {(31,0.0) (31,0.5)};
            
            \addplot[name path=reuna6, mark=none, draw=none] coordinates {(28.5,1.9) (28.5,2)};
            \addplot[name path=reuna7, mark=none, draw=none] coordinates {(31,1.9) (31,2)};
            
            \addplot [gray!30] fill between [of = reuna4 and reuna5,    reverse=true];
            \addplot [gray!30] fill between [of = reuna6 and reuna7,    reverse=true];

            \addplot[name path=hantar,  mark=none, dashed] coordinates {(30,0.5) (30,1.9)};
            
            \addplot[name path=loppur,  mark=none, draw=none]  coordinates {(35,0.5) (35,1.9)};
            
            \addplot [yellow!60] fill between [of = hantar and loppur,    reverse=true];
            
            \addplot[name path=kevyt1,  mark=none, dashed] coordinates {(29.6,0.5) (29.6,1.9)};              
            
            \addplot[name path=hantak,  mark=none, dashed] coordinates {(30,0.5) (30,1.9)};
            
            \addplot [red!30] fill between [of = hantak and kevyt1,    reverse=true];


            %
        \end{groupplot}
        
        \draw(dl)--++(0.1,0)--+(0,0.25)--+(0,-0.25);
        \draw(dr)--++(-0.1,0)--+(0,0.25)--+(0,-0.25);


            %
            
            
    \end{tikzpicture}               
\end{subfigure}
\caption{Regions of the hybrid interpolation method for approximating symmetric $\alpha$-stable laws. 
    Turquoise: the first bicubic interpolation grid. Violet: the second bicubic interpolation grid. Red: the transition region of the spline and the asymptotic series. Yellow: the asymptotic series expansion  for $r \rightarrow \infty$. Gray: the implemented approximation method is not employed.}
\label{fig:system}
\end{figure}

To address the issues of the existing approximation methods in Bayesian inversion, our hybrid approximation method is divided into parts, and different techniques are employed within them. When $r$ is small, we approximate the $\alpha$-stable laws with two-variable bicubic splines that are fitted on  grids of precomputed $\alpha$-stable log densities with varying radius $r$ and stability $\alpha$. To our knowledge, a similar rationale has been presented by \citet{MR06}, but our  method  works also for $\alpha < 1$. Furthermore, our method approximates the probability densities on a continuous range of the stability indices $\alpha$ with consistent accuracy and regularity, and for a continuous range of $\alpha$. The feature is essential for hierarchical $\alpha$-stable prior constructions.   We employ the  Julia library  \texttt{Interpolations.jl} to evaluate the bicubic splines.  Figure \ref{fig:system} depicts the overall approximation method. We present the details of the parts of the hybrid method below.

\subsubsection{The first bicubic spline approximation}

The first bicubic spline grid of precomputed log densities is applied when  $r \in [0,0.9], \, \alpha \in [0.5, 1.9] $. This region is illustrated in turquoise in Figure~\ref{fig:system}. We divide the domain $[0,1.0]\times [0.5,1.9]$ uniformly to the intervals of $h_{r} = 0.01 $ and $ h_{\alpha}=5\cdot10^{-4}$, and evaluate numerically the densities using the integral methods. For the one-dimensional $\alpha$-stable density we use the Fourier integral  formula from \eref{direct} for $|\alpha - 1| < 0.2$, and otherwise we employ Nolan's method of \eref{nolanintegral}. 
Instead of introducing improvised heuristics for the integration and domain partitioning, we count on the Fourier integral for the aforementioned stability values that are particularly tricky in Nolan's method. 
The numerically evaluated densities of both methods agree each with a least 12 decimals for $|\alpha -1| \geq 0.2$, so using only the Fourier approach would suffice. 

We use the very same spline grid  for the bivariate symmetric $\alpha$-stable laws, and we employ the integral expression with the Bessel function in \eref{multivar}. In both cases, we set the derivative with respect to $r$ at $r=0$  to zero in the spline to ensure the method approximates the partial derivative with respect to $r$ properly at $r=0$. At $r=0.9$, we set the partial derivatives  with respect to $r$ to agree with those given by the second spline, which guarantees the  $C^1$ continuity of the spline approximation for the log density at $r=0.9$, which is the upper limit of the first spline.

\subsubsection{The second bicubic spline approximation}
The second bicubic spline grid is constructed in  $r \in [0,30], \, \alpha \in [0.5, 1.9]$ with the intervals of $h_{r} = 0.01 $ and $ h_{\alpha}=5\cdot10^{-4}$ for the nodes of the grid. The domain is illustrated in violet in Figure~\ref{fig:system}. We apply the same approach as in the first grid For the one-dimensional $\alpha$-stable laws, we use Nolan's method for approximating the log densities in the grid points $|\alpha-1| > 0.2$, and otherwise we use the inverse Fourier method. Likewise, we use the integral expression incorporating a Bessel function for the bivariate $\alpha$-stable laws to construct the log density grid. 

We apply natural boundary conditions with respect to $r$ and $\alpha$ in the spline, that is, the second derivatives with respect to them are set to zero on the boundaries of the spline. However, we limit the usage of the spline only for $r \in (0.9,29.6), \, \alpha \in [0.5, 1.9]$.  Again, we want the overall approximation and its partial derivatives to be continuous. That is why we intentionally let this second spline grid, which is also the largest of them all, overlap with the domains of the first and the third spline grids, and modify their boundary conditions to match the partial derivatives with respect to $r$ of the second spline. 

\subsubsection{Series expansions for approximating the tails}
As we pointed out in the overview, approximating the $\alpha$-stable laws for $r\geq 0$ is ineffective with numerical integration-based approaches, so we use the asymptotic series expansions for the tails of the distributions. For the univariate $\alpha$-stable log densities, we use  \eref{tail}, and for the bivariate tails, we use \eref{multiseriesinfty}.  We use three terms in both series expansions, which we found to provide feasible accurate tail density approximations.  The domain of the series expansions is depicted in yellow in Figure~\ref{fig:system}. 

The lower limit of $r=30$ for the series expansions was selected because it has been reported to provide a practical rounded  bound for the asymptotic tail expansions of the univariate stable densities up to $\alpha=1.999$. However, the number of required terms in the series can be several tens or over a hundred for very high accuracy  \cite{TA08}.  Naturally, such a high number of terms would result in a loss of computational efficiency. We keep the same lower limit for the tail expansions of the symmetric bivariate $\alpha$-stable distributions since we have numerically verified the difference of the log densities given by the tail approximation and the numerical integration-based approach  are of the same magnitude as in the univariate case. 

\subsubsection{The transition region}
The  switch from the spline grids to the asymptotic tails series is made seamlessly with the help of  the third bicubic spline. The spline is employed for $r \in [29.6,30], \, \alpha \in [0.5, 1.9]$, and  we call this region the transition region of the approximation. The purpose of the transition region is to ensure the log-density approximation and its partial derivatives are continuous everywhere in its domain.   The domain of the transition region spline is illustrated in red in Figure~\ref{fig:system}. We use the same grid node intervals of $h_r = 0.01 $ and $ h_{\alpha}=5\cdot10^{-4}$ in this domain as we do in the other two grids, and the methodology is identical for both in the univariate and    bivariate $\alpha$-stable laws.
The spline is constructed as follows. Let us denote  $r_a = 29.6$, $r_b = 30.0$, and $\Delta= r_b- r_a$. Let the log-density approximation given by the asymptotic series expansion from  \eref{tail} (\eref{multiseriesinfty} in the bivariate case) at the spline grid node  $r_a,\alpha_j$ be $f^s_{a,j}$ and its derivative with respect to $r$ be  $D^s_{a,j}$. Additionally, let $f^d_{b,j}$ stand for the log-density approximation given by the numerical integration at $r_b,\alpha_j$, and  $D^d_{b,j}$ denote its derivative with respect to $r$. We set the value in the transition region spline grid point value at $i,j$  to follow an auxiliary cubic Hermite interpolation as follows:
\begin{align}
\label{muokkaus}
    f_{i,j} =& \frac{(3\Delta q_i^2-2q_i^3)}{\Delta^3} f^s_{b,j} + \frac{(\Delta^3-3\Delta  q_i^2+2q_i^3)}{\Delta^3} f^d_{a,j} + \frac{q_i^2 (q_i-\Delta)}{\Delta^2}D^d_{b,j}  \cr &+ \frac{q_i(q_i-\Delta)^2}{\Delta^2}D^s_{a,j},
\end{align}
where  $q_i = r_i - r_a$. 

Equation \eref{muokkaus} is applied for each stability $\alpha_j$ within the transition grid separately. The Hermite interpolation is only applied during the construction of the transition spline because the evaluation of the constructed grid is performed by the bicubic spline library in Julia.  Introducing the Hermite interpolated data points as an additional step at the transition region helps to avoid abrupt changes in the derivatives of the log densities near the boundary of the transition region and the tail approximation. 

The derivatives of the spline with respect to $r$ on the boundary $r=29.6$ are set to follow the values given by the second spline. The same derivative at $r=30$ is set to follow the derivative given by the asymptotic tail expansion, respectively.  

\subsection{Error bounds of the hybrid method}

We obtain the following  error bound for the symmetric $\alpha$-stable log densities in the domain of the spline grids:
\[\fl
\sup_{r \leq 30, \; 0.5 \leq \alpha \leq 1.9} \,
|\log\,f_T(r;\alpha) - \log\, f_A(r;\alpha)| \leq
\begin{cases}
   0.00038  & \quad \textrm{(univariate case);} \\
   0.22  & \quad \textrm{(bivariate case),}
\end{cases}
\]
where   $f_T \defeq f(r;\alpha)$ stands for the (true) density of the symmetric $\alpha$-stable distribution with $\sigma = 1$, and $f_A \defeq f_A(r;\alpha)$ denotes density given by the bicubic spline interpolation.
%
%
The error estimate for the bivariate log density is  orders of magnitudes higher than for the univariate one due to the significantly larger suprema for the partial derivatives within the domain of the splines, particularly with small $\alpha$. If the lower bound of $\alpha$ of the approximation domain was increased to 0.7, the bivariate log-density error estimate would  decrease to 0.013, accordingly. The accuracy of the approximations is enough for our needs in the Bayesian inversion.

For the relative error bounds of the tails, we have the following estimates. Denoting by $\mathcal{S}_3(r;\alpha)$ the sum in \eref{tail} (resp.~\eref{multiseriesinfty}) with $3$ in place of $\infty$, and by $f_T(r;\alpha)$ the true density, we have
\[\fl
\sup_{r > 30, \; 0.5 \leq \alpha \leq 1.9} \,
|\log\,f_T(r;\alpha) - \log\, \mathcal{S}_3(r;\alpha)| \leq
\begin{cases}
0.00097  & \quad \textrm{(univariate case);} \\
0.0017   & \quad \textrm{(bivariate case).}
\end{cases}
\]

For clarity, we briefly explain the rationale of the procedures behind the derived bounds. First, we assume the integration error of the probability densities to be zero within the spline grid points. For the second, we ignore the transition region spline from the error estimates, because it can be  left out of the hybrid method at the expense of having a less regular overall approximation. Then we make use of the properties of bicubic splines as follows \cite{HW76}. 
\begin{theorem}
Let the true $\alpha$-stable density and the bicubic spline approximated density be denoted as above.  The error resulting from the bicubic spline approximation can then be approximated by
\cite{CH73,HW76}
\begin{align}\label{eq:error}
        \fl \Vert \log f_T - \log f_A\Vert_{\infty} \leq & \frac{5}{384}\Vert(\log f_T)^{(4,0)}\Vert_{\infty} h_r^4 +  \frac{81}{64}\Vert(\log f_T)^{(2,2)}\Vert_{\infty} h_r^2 h_{\alpha}^2 \\&
        + \frac{5}{384}\Vert(\log f_T)^{(0,4)}\Vert_{\infty} h_{\alpha}^4,
\end{align}
where $h_r$ and $h_{\alpha}$ stand for the intervals of the $\log$ density interpolation grid cells in the directions of the radius and the stability index, respectively, and the superscripts $^{(i,j)}$ stand for partial derivatives of the form $\frac{\partial^{i+j}}{\partial r^i \partial \alpha^j}$.
\end{theorem}

Due to the lack of any sort of a closed-form expression for $f_T$, estimating the partial derivatives of $\log f_T$ appearing in the suprema in \eref{eq:error} involves estimating the partial derivatives of $f_T$ from above, and $f_T$ itself from below. 
Both types of estimates are tricky to do in a precise manner.
When estimating the partial derivatives of $f_T$, we will use several strategies that are variably efficient for different regions of $(r;\alpha)$, and then use the first-order variants of these estimates in conjunction with a precomputed grid 
and the fundamental theorem of calculus for the lower bounds of $f_T$. The full details of these estimates are presented in the Supplementary Material.

First, we can obtain crude uniform bounds (with respect to $r$) for each $(f_T)^{(i,j)}$ by e.g.~differentiating \eref{direct} under the integral sign and eliminating the resulting oscillatory term simply using the triangle inequality. We can somewhat refine this pointwise for ``moderate'' $r > 1$ by using standard oscillatory integral techniques (which basically amount to partial integration against a sufficiently quickly vanishing function, resulting in an upper bound of order $r^{-1}$). 
This is largely the best we can do for moderate values of $r$, where neither of the asymptotic expansions (see \eref{tail} and the subsequent discussion) come close to approximating $f_T$ well with only a couple summands -- the region of said ``moderate'' values will, of course, depend on $\alpha$.

For larger values of $r$, we may exploit the expansion \eref{tail} pointwise with e.g.~$2$--$3$ summands and an explicit (albeit complicated) expression for the remainder term, due to Bergstr\"om \cite{HB52}. Some of the integrals we encounter here 
are highly intractable in the mathematical sense of the word, but non-oscillating and well-behaved enough for efficient numerical estimation, yielding an upper bound that decreases to order roughly comparable to that of $f_T$ for $r \to \infty$, and where the asymptotic constants stay sufficiently tame for $\alpha \in [0.5,1.9]$. This all holds, mutatis mutandis, for $r \to 0^{+}$ as well.

As a result, we obtain several different kinds of upper bounds for $|(f_T)^{(i,j)}(r;\alpha)|$, pointwise with respect to $(r,\alpha)$. With some minor additional work, we may loosen the estimates very slightly so that they will be uniform for the spline grid cells $r \in [r_j,r_j + h_r]$ and $\alpha \in [\alpha_i,\alpha_{i} + h_\{\alpha]$, where $i$ and $j$ are indexed over the numbers of the spline grid points.
%
With the upper estimates for the partial derivatives of $f_T$, we then estimate $f_T$ from below with reasonable accuracy. Namely, noting that $f_T(r;\alpha)$ is for fixed $\alpha$ always a decreasing function of $r$, we may precompute $f_T(r_j;\alpha_i)$ at the nodes of the discretized grid, and thus use the fundamental theorem of calculus to obtain the lower bound for $f_T(r_j;\alpha_i)$ within the spline grid cells.

The discussion above also applies to the bivariate case, with the additional difficulty of the Bessel function of the first kind $J_0$ in the representation \eref{multivar}. In the Supplementary Material, we present analogous asymptotic expansions with respect to $r \to 0^{+}$ and $r \to \infty$ with quantitative remainder term estimates. In particular for $r \to \infty$, we present a modification of Bergstr\"om's \cite{HB52} complex-analytical treatment, which allows us to obtain estimates for the remainder term in the bivariate case which are not immediate from the asymptotic expansions presented in \cite{JPN13}. 

\subsection{Practical considerations on the hybrid method}

We consider the computational efficiency of the presented hybrid scheme as good. The evaluation of an $\alpha$-stable log density takes approximately 100 nanoseconds  in the domain of the spline grids, and about 400 nanoseconds in the asymptotic tail expansions on a workstation equipped with Intel Xeon CPU E5-2698 v4 central processing unit. For instance, evaluation of the univariate $\alpha$-stable density through Nolan's method or the inverse Fourier transform approach with a relative tolerance of $10^{-10}$ takes approximately 10-150 microseconds for $\alpha= 1.6\, r \in [0,30]$ using Julia's \texttt{QuadGK.jl} library, which applies adaptive Gauss–Kronrod quadrature. However, for $\alpha= 1.6\, r = 150$, both integration-based methods take over 2 milliseconds to evaluate, and the computational efficiency degrades the further the greater the radius argument $r$ is because the integrals are close to zero, and hence slower to evaluate with the specified relative tolerance. On a typical Bayesian inverse problem, the number of separate evaluations of probability density functions can be tens or hundreds of thousands in a posterior distribution. Furthermore, an approximation based on combining power series expansions such as the method of \citet{CAMM18} is likely not any faster than the hybrid method that combines splines with well-known asymptotic tail expansions since the method requires evaluation of two different series expansions simultaneously. That is why a hybrid method is essential from a computational efficiency viewpoint, and we argue that our method certainly meets those requirements. 

In the presented methodology, we do not consider the stability indices of  $\alpha < 0.5$ or $\alpha > 1.9$.  In theory, values of $\alpha$ close to $2-10^{-6}$ can be effectively approximated with the presented hybrid method, because the series approximation and the integration-based spline interpolation agree well at $r=30$ for that high $\alpha$. However, it must be borne in mind that the closer the stability index is to 2, the worse the relative error of the approximation will be because of the magnitudes of the partial derivatives with respect to $\alpha$. As a remedy to that, an extra spline grid could be incorporated in the approximation for $1.9\leq\alpha\leq 2-10^{-6}$, but the problem in the approach is to smoothly join the splines through their boundary conditions with respect to $\alpha$, which would further complicate the overall workflow.  
On the other hand, low stability indices are not in our interest, and including them would require increasing the number of nodes in the precomputed log-density  grids to sustain the accuracy of the approximation. Alternatively, yet another spline grid could be employed for smaller $\alpha$ with a possibly greater range of $r$, which would again make the method more complex. We remark that the inverse Fourier transform-based method copes badly at approximating the univariate $\alpha$-stable densities with a small stability index, whereas Nolan's method still sustains its effectiveness thanks to the compact integrand it adapts.

Finally, we elaborate on the selection of the spline parameters.  The first and the transition region splines ($r\in [0,0.9]$ and $r\in [29.6,30.0]$) are useful also because the resulting systems of equations of the spline coefficients involving the boundary conditions of the splines are smaller, and hence easier to solve than directly incorporating them into the coefficients of the largest spline grid. That is because the interpolation library \texttt{Interpolations.jl} applies effectively the symmetries of the spline coefficient equations when natural boundary conditions are used.  The limit of $r=29.6$ for the transition region splines was selected as a good compromise between the regularity of the approximation and its relative accuracy. 
The limit of the first spline ($r=0.9$) can be argued to be adequate for the first spline because the partial derivative with respect to $r$ is the largest with the small $\alpha$, and the largest magnitudes are obtained for $\alpha < 0.9$. We have not optimized the limit further because all spline grids have the same node intervals. However, it would be straightforward to decrease the spacing  $h_r$ in the first spline, if the lower stability indices will be of interest in the future. As a proof of concept, we regard the current grid intervals of $h_r=0.01$ and $ h_{\alpha}=5\cdot10^{-4}$ as a practical compromise between accuracy and the required memory to store grids. The current parameters result in a grid of approximately $3000\times3000$ nodes in the second interpolation spline. Although all the grids in the study need only a few hundred megabytes of storage in total, we emphasize that the asymptotic series expansions are virtually the largest source of the relative error in the overall approximation despite the proven error bounds. Thus, we consider decreasing the node intervals impractical for now.





\section{Numerical examples}\label{sec:num}
We demonstrate the $\alpha$-stable  priors in three numerical experiments. 
We employ the priors first in a deconvolution, which is a well-known linear inverse problem. Moreover, the same priors are used in estimating the conductivity field of an elliptic partial differential equation in two spatial dimensions. 
For the time being, we only use MAP estimators in the reconstructions because full Bayesian inference with the presented random field priors requires the usage of MCMC methods, which have been shown to struggle with such heavy-tailed priors \cite{SCLR21}. Since the assessment of the reconstructions in inverse problems cannot be usually accomplished in a unified manner, we do not intentionally tabulate any metrics of the reconstructions, such as $L^2$ errors of the reconstructions, in the manuscript. We demonstrate the MAP estimators of the $\alpha$-stable priors by varying the stability index $\alpha$ and the scale $\sigma$ in the examples. We did not optimize their ranges in the experiments. Rather, they were selected so that their effect on the estimators would be meaningful to illustrate.  The Julia codes of the experiments can be found at
\url{https://github.com/suurj/alpha-stable}.

\subsection{MAP estimation}

Evaluation of the MAP estimators \eref{mappi} in Bayesian continuous-parameter estimation is usually performed with the help of a nonlinear conjugate gradient algorithm, a quasi-Newton method, a matrix-free truncated Newton method, or a combination of them \cite{ANS09, IL14, MBOV13,YRC19}.
%
%

In our numerical examples, the  maximization of the  log posteriors is done through the L-BFGS method in the deconvolution experiments. 
Moreover, we resort to the bounded L-BFGS algorithm \cite{ZBLN97} at the inversion of the conductivity field of a linear elliptic partial differential equation. 
As the numerical implementations of the limited-memory BFGS algorithms, we use \texttt{Optim.jl} \cite{Optim} for the unconstrained L-BFGS, and a Julia wrapper \texttt{LBFGSB.jl}  of the original Fortran-implementation of L-BGFS-B \cite{ZBLN97}. 
Lastly, we  want to emphasize that the presented $\alpha$-stable random field priors often make the posteriors multimodal \cite{SCLR21}, and finding global maxima from such distributions is difficult. Moreover, different optimization algorithms may converge to different local minima. 

\subsection{One-dimensional deconvolution}

\begin{figure}
\centering
\begin{subfigure}[t]{5.5cm}
    
    \includegraphics[scale=0.4]{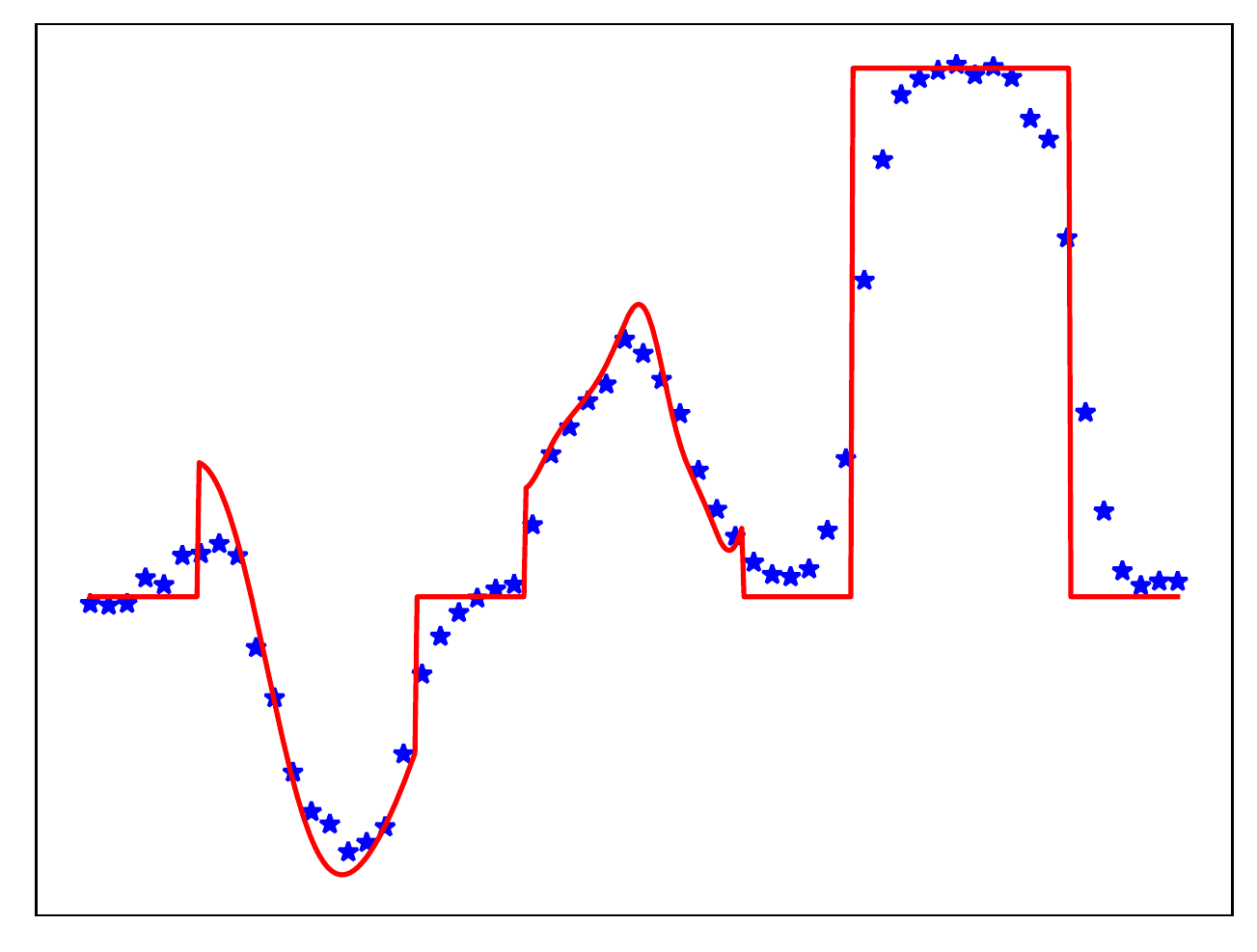}
    \caption{Ground truth function (red) and noisy convolutions (blue).}
    \label{fig:gt}    
\end{subfigure}

\centering   
\includegraphics[width=1.0\linewidth]{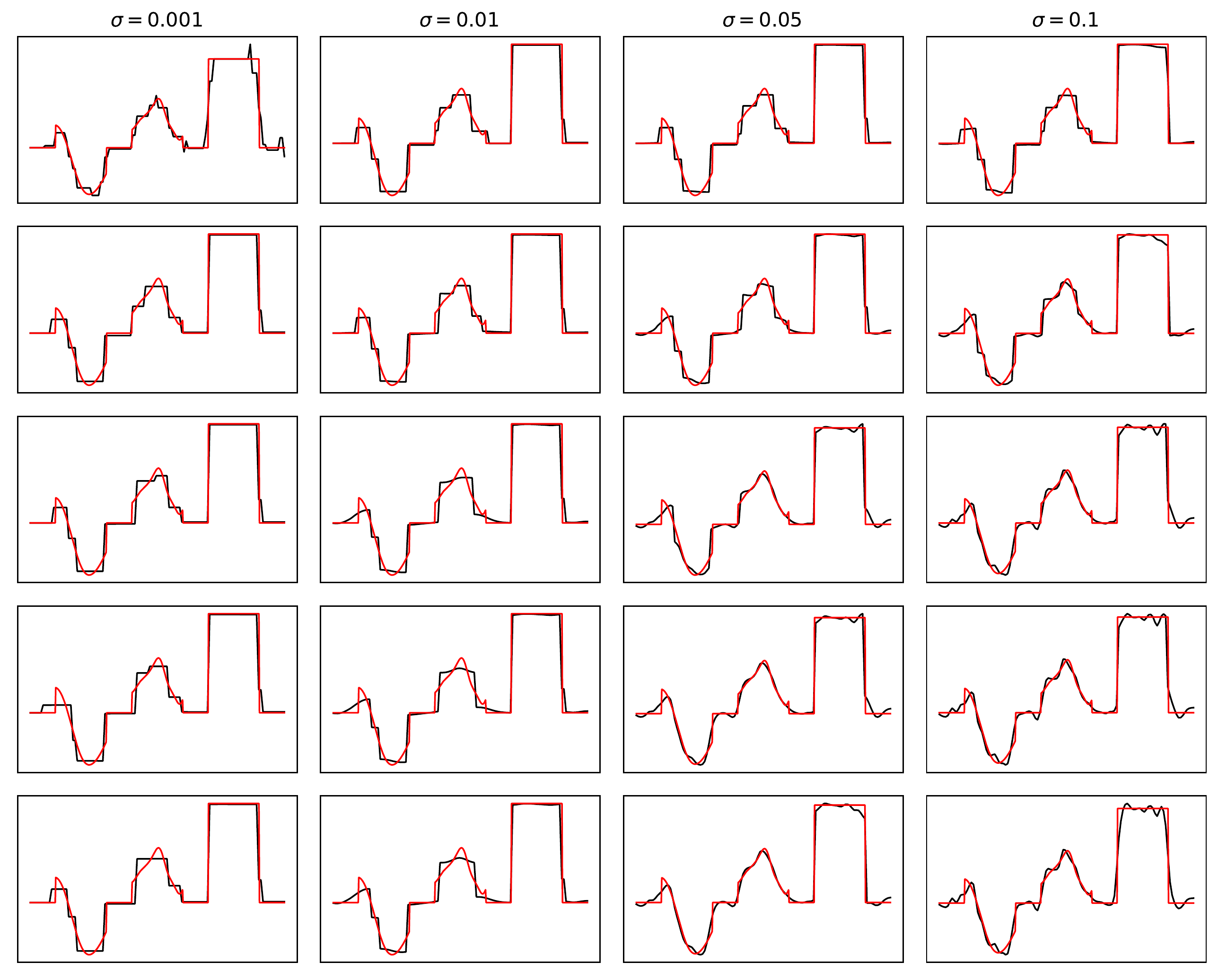}   

\caption{Ground truth, measurements, and the reconstructions with fixed stability index and scale parameter in the one-dimensional deconvolution experiment with $\alpha$-stable difference prior. Red lines: ground truth. Black lines: MAP estimate for the function.  }
\label{plain}
\end{figure}

In the first numerical experiment, we demonstrate the properties of the first-order $\alpha$-stable difference priors. We discretize the target function $u$, which includes both discontinuities and piecewise realizations of a Gaussian process with Matérn covariance, at 500 equispaced grid points on $[-1,1]$. We convolve $\mathbf{u}$  with a normalized Gaussian kernel of
\begin{equation}
\label{kerneli}
\psi(p) = 25 \exp\left(-50 \Vert p \Vert^2\right)
\end{equation}
though a matrix approximation, which is denoted by $\mathbf{F}$. 
We construct the noise-corrupted data $\mathbf{y}$ at 60 equispaced points within the support of the target function by adding white Gaussian noise to the discrete convolution as follows:
\begin{equation}
\label{suoramalli}
\mathbf{y} =  \mathbf{F}\mathbf{u}    + \bm{\epsilon}, \, \bm{\epsilon} \sim \mathcal{N}(\bm{0},0.02^2\mathbf{I}). 
\end{equation}
Thus, the likelihood $\pi(\mathbf{y}\mid\mathbf{u})$ of the forward model is Gaussian. 
We avoid committing an inverse crime by reconstructing $\mathbf{u}$ on a 120-dimensional equispaced grid at $[-1,1]$ and use the corresponding matrix approximation for the convolution with the kernel \eref{kerneli} on the reconstruction grid to evaluate the forward model and the likelihood. The ground truth function $u$ and the measurement data $\mathbf{y}$ are depicted in Figure \ref{fig:gt}.  

\begin{figure}
\centering
\includegraphics[width=1.0\textwidth]{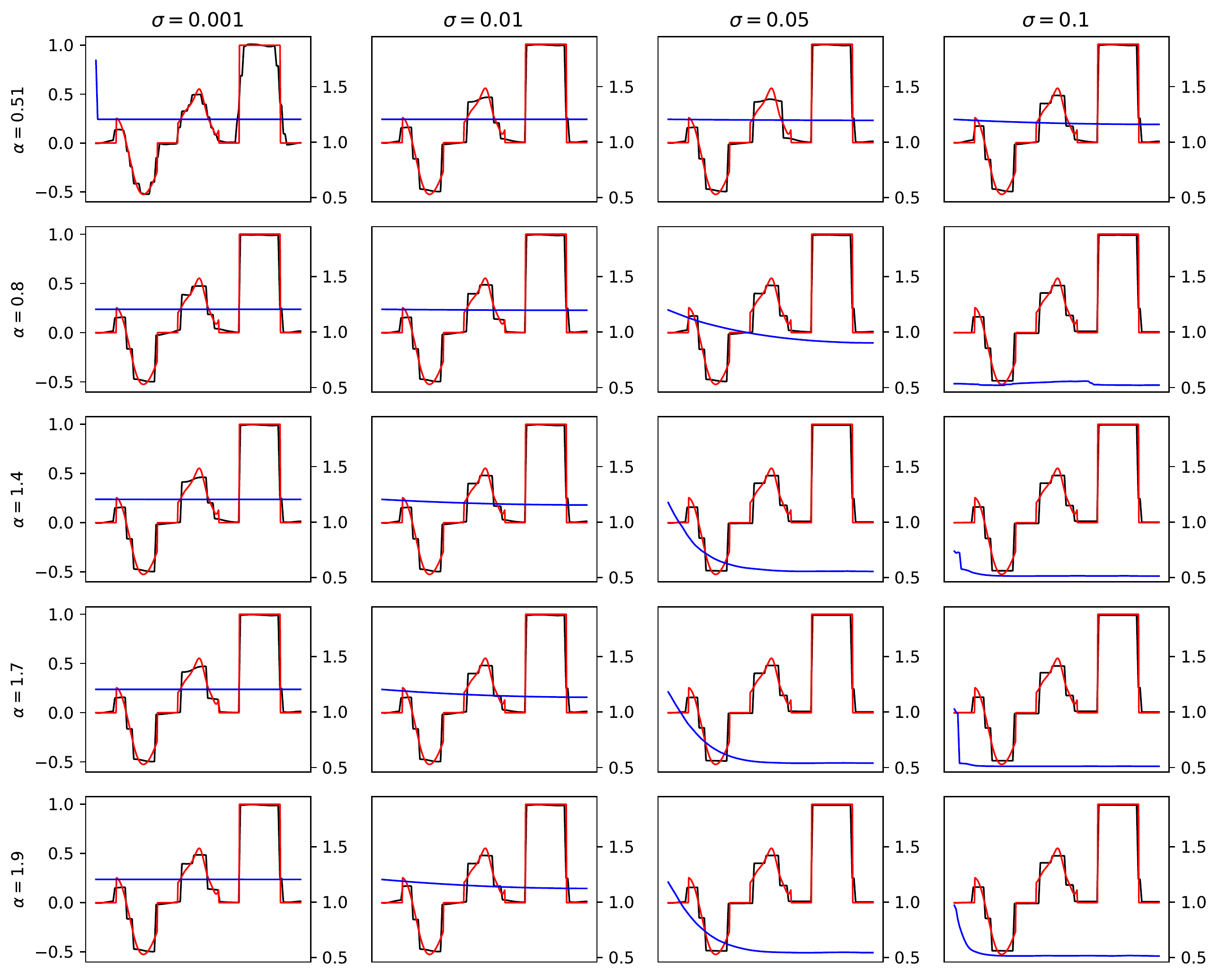}
\caption{Reconstructions when considering the stability $\alpha$ as a process. Red lines: ground truth. Black lines: MAP estimate for the function. Blue lines: MAP estimate of the stability process.  }
\label{stabi}
\end{figure}

\begin{figure}
\centering
\includegraphics[width=1.0\textwidth]{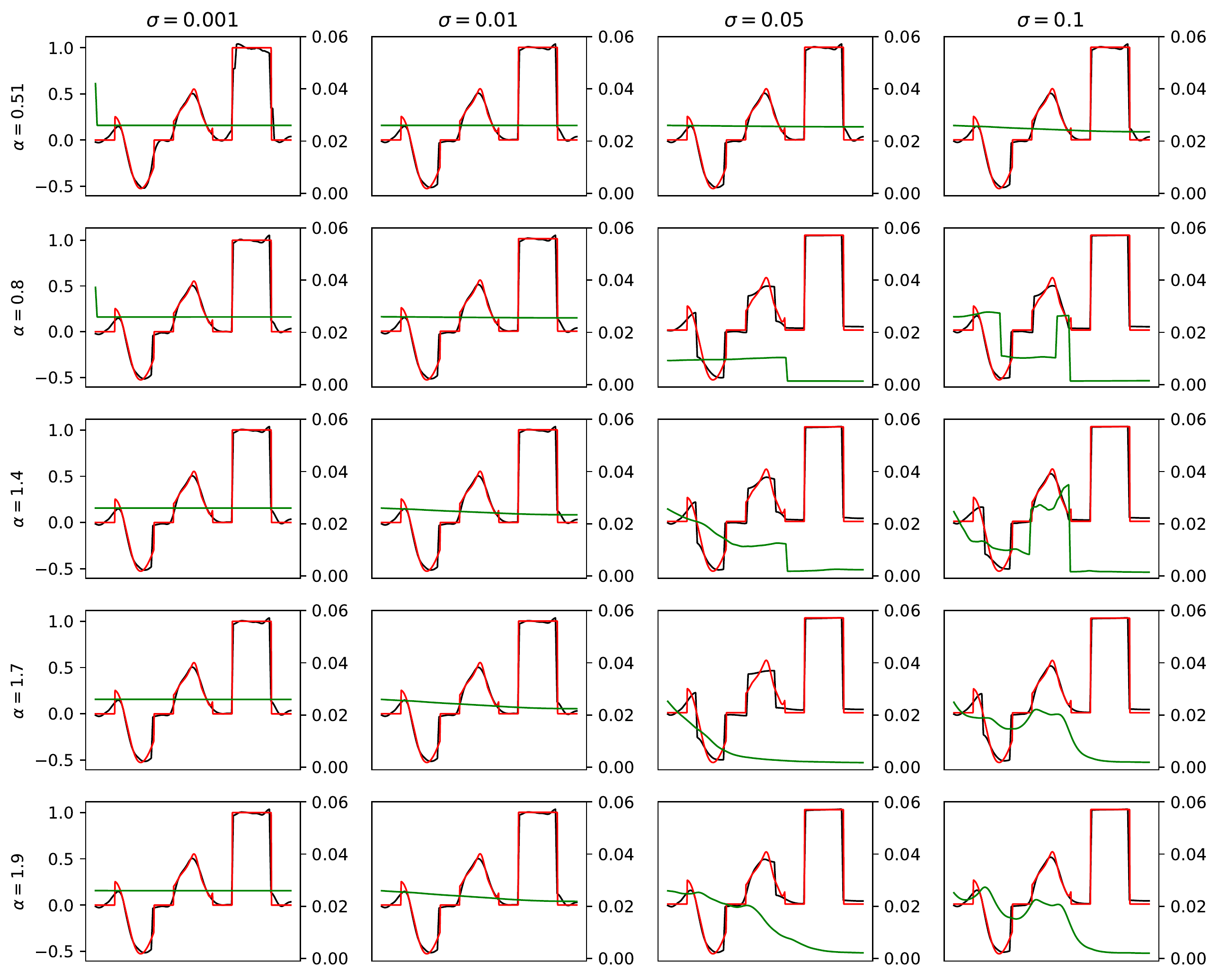}
\caption{Reconstructions when considering the scale $\sigma$ as a process. Red lines: ground truth. Black lines: MAP estimate for the function.  Green lines: MAP estimate of the scale process.  }
\label{scale}
\end{figure}

\begin{figure}
\centering
\includegraphics[width=1.0\textwidth]{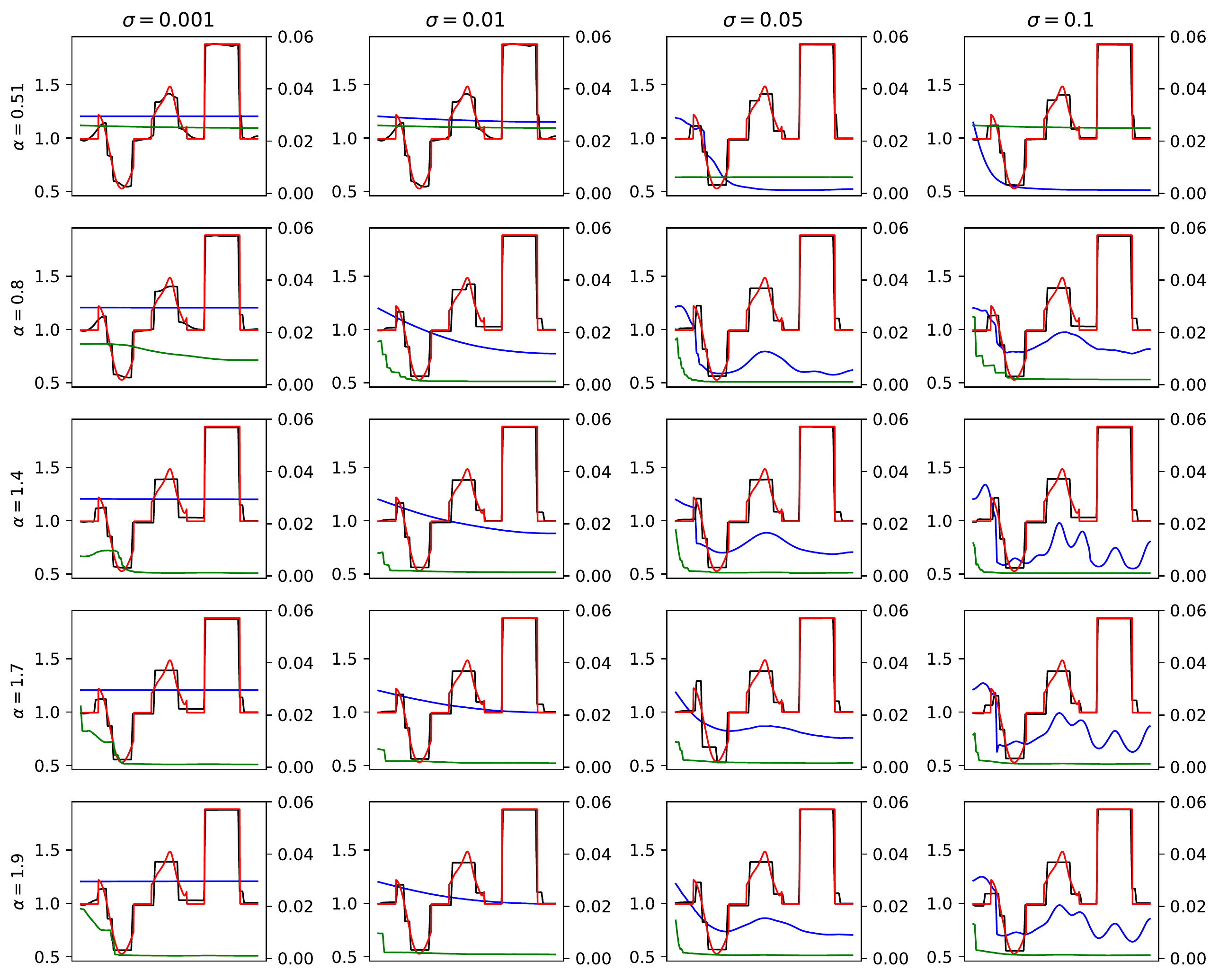}
\caption{Reconstructions when considering both the stability $\alpha$ and the scale $\sigma$ as  processes. Red lines: ground truth. Black lines: MAP estimate for the function. Blue lines: MAP estimate of the stability process. Green lines: MAP estimate of the scale process.   The left axes of the subfigures stand for the stability indices, and the right ones denote the scale.}
\label{both}
\end{figure}

We employ four different $\alpha$-stable priors in the one-dimensional setting. Namely, we exploit the hierarchical $\alpha$-stable priors defined in  \eref{hiera}, so that
\begin{enumerate}
\item fix the scale $\sigma$ and stability index $\alpha$ of the prior of the increments of $\mathbf{u}$,
\item consider the stability index  of the prior of the increments of $\mathbf{u}$ as a process that depends on another $\alpha$-process  $\mathbf{s}$ and fix the scale of the increments,
\item consider the scale of the prior of the increments of $\mathbf{u}$ as a process that depends on another $\alpha$-process $\mathbf{c}$ and fix the stability index of the increments,
\item and consider both the stability index and the scale of the prior of the increments of $\mathbf{u}$ as processes that depend on another $\alpha$-processes $\mathbf{s}$ and $\mathbf{c}$. 
\label{lista}
\end{enumerate}
The MAP estimates with the priors are illustrated in Figures \ref{plain}, \ref{stabi}, \ref{scale}, and \ref{both}. The selected discretization of the processes implies that the dimensions of the posterior distributions are 120-, 240-, 240-, and 360-dimensional, respectively.  

The MAP estimates with the non-hierarchical $\alpha$-stable first-order difference priors in Figure \ref{plain} demonstrate the effect of altering the stability  $\alpha$ or scale $\sigma$ of the distribution of the increments in the  prior of $\mathbf{u}$. As a rule of thumb, the smaller the stability $\alpha$ is, the stronger the prior favors non-Gaussian increments, so they are usually close to zero. The larger the scale $\sigma$ is, the greater the variability is allowed within the increments. Stability indices of $ 1 \leq \alpha \leq 2$, are particularly useful for reconstructing the target function in this example case. Those priors are able to favor the existence of Gaussian-like parts of the ground truth function when needed. If the estimation was done using stationary Gaussian priors, the MAP estimate would be either over-smoothed and incapable to locate the discontinuity at the boxcar, or it could detect the discontinuity at the expense of being very sensitive to noise.

Considering the stability of the prior of $\mathbf{u}$ as another first-order $\alpha$-stable process, turned out to be less successful. We let the scale of the process $\mathbf{u}$ to follow an $\alpha$-stable process with scale $\sigma=0.01$ and set its untransformed stability process $\mathbf{s}$ to follow an $\alpha$-stable process with the parameters tabulated in Figure \ref{stabi}. To guarantee that  $0.51 \leq \alpha \leq 1.9$, we apply a  transformation \eref{hiera}
\begin{equation}\label{stabtr}
\alpha = H(s) =  0.51 + 1.39 S(s),
\end{equation}
where $S(x) = \frac{1}{1+ \exp(-x)}$. In both Figures \ref{stabi} and \ref{both}, the stability processes are  shown in their transformed values. The stability process seems to be either close to constant ($ \approx 1.25$) or decreasing towards the right side of the domain  in all the tabulated cases. 
However, there is some variation in the stability process in the middle of the domain when the untransformed process has the parameters $\alpha=0.8, \sigma=0.1$. The phenomenon may suggest that the stability index being a process  does not work well as a prior. 
When the stability  of the untransformed stability process is $\alpha_s=1.4$ and its scale $\sigma_s=0.05$ (Equation \eref{hiera}), the reconstruction of $\mathbf{u}$ is smooth at first, but as the stability decreases, the reconstruction becomes more discontinuous and non-Gaussian. 

In the one-dimensional deconvolution experiment, the best results in terms of the reconstruction agreement with the ground truth are obtained when the scale of $\mathbf{u}$ process is considered  a process instead of its stability. 
We fix the stability of $\mathbf{u}$  to $\alpha=1.9$ and instead model the untransformed discretized scale process $\mathbf{c}$ with another $\alpha$-stable process with the parameters shown in Figure \ref{scale}. The final scale process  \eref{hiera} is given by a transformation 
\begin{equation}\label{scaletr}
\sigma = G(c) = 0.001 + 0.05 S(c).
\end{equation} 
The reconstructions where the untransformed scale  process has the scale of $0.05 \leq \sigma_c \leq 0.1$ and stability between $0.8\leq \alpha_c \leq 1.9$, agree well with the ground truth and even with each other.

For the last, setting both the scale and the stability  of $\mathbf{u}$ as $\alpha$-stable processes seem to suffer from the same issue as the stability process case.
We set the scale of the untransformed stability index process to $\sigma_s=0.05$ (Equation \eref{hiera}), and the stability of the untransformed scale parameter process to $\alpha_c=1.9$. Hence, the scale parameters $\sigma$ in Figure \ref{both} refer to the scale of the untransformed stability process ($\sigma_s$), and the the stability indices to the  untransformed scale process ($\alpha_c$). We transform the processes with the same sigmoid functions as in the other two cases, using Equations \eref{scaletr} and \eref{stabtr}. Either one or both of the parameter processes remain close to constant throughout the domain, and the MAP estimates for $\mathbf{u}$ are no better than in the simpler hierarchical $\alpha$-stable priors.    Whether the poor reconstructions are caused by overfitting, poorly selected hyperparameters of the processes $\bm{s}$ and $\bm{c}$, unidentifiability, or something else, shall be investigated in further studies.

\newcommand{\qhei}{3.4cm}
\newcommand{\phei}{3.4cm}
\newcommand{\boxi}{1pt}
\newcommand{\nosto}{2pt}
\newcommand{\boxip}{5pt}
\newcommand{\nostop}{1.4cm}
\newcommand{\nostoq}{0.2cm}
\newcommand{\nostor}{1.1cm}
\newcommand{\nostos}{0.2cm}
\newcommand{\nostot}{0.2cm}

\subsection{Two-dimensional deconvolution}

\begin{figure}
\centering
\begin{subfigure}[t]{5.4cm}
    \centering
    \includegraphics[width=3.4cm]{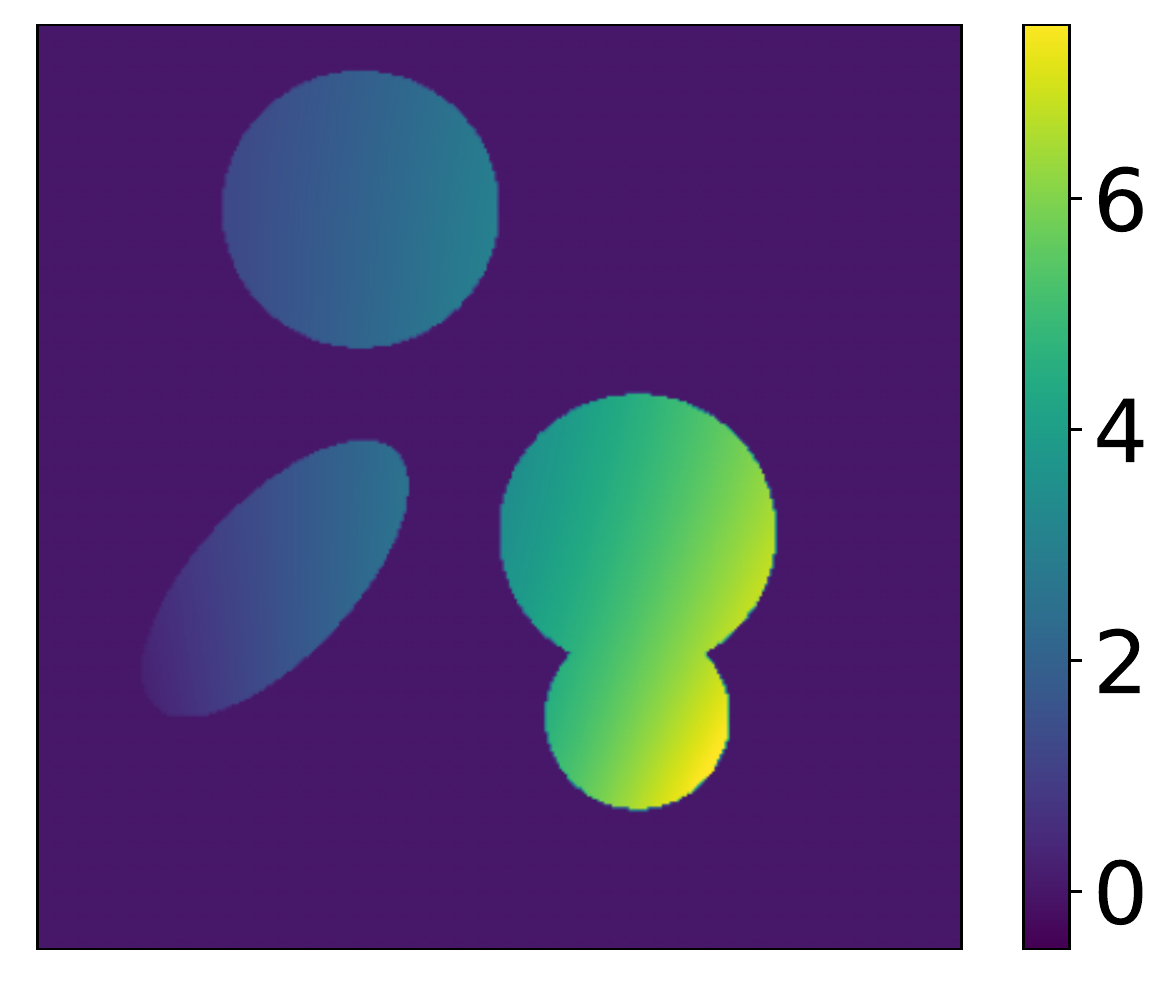}    
    \caption{Ground truth function.}  
\end{subfigure}\begin{subfigure}[t]{5.4cm}
    \centering
    \includegraphics[width=3.4cm]{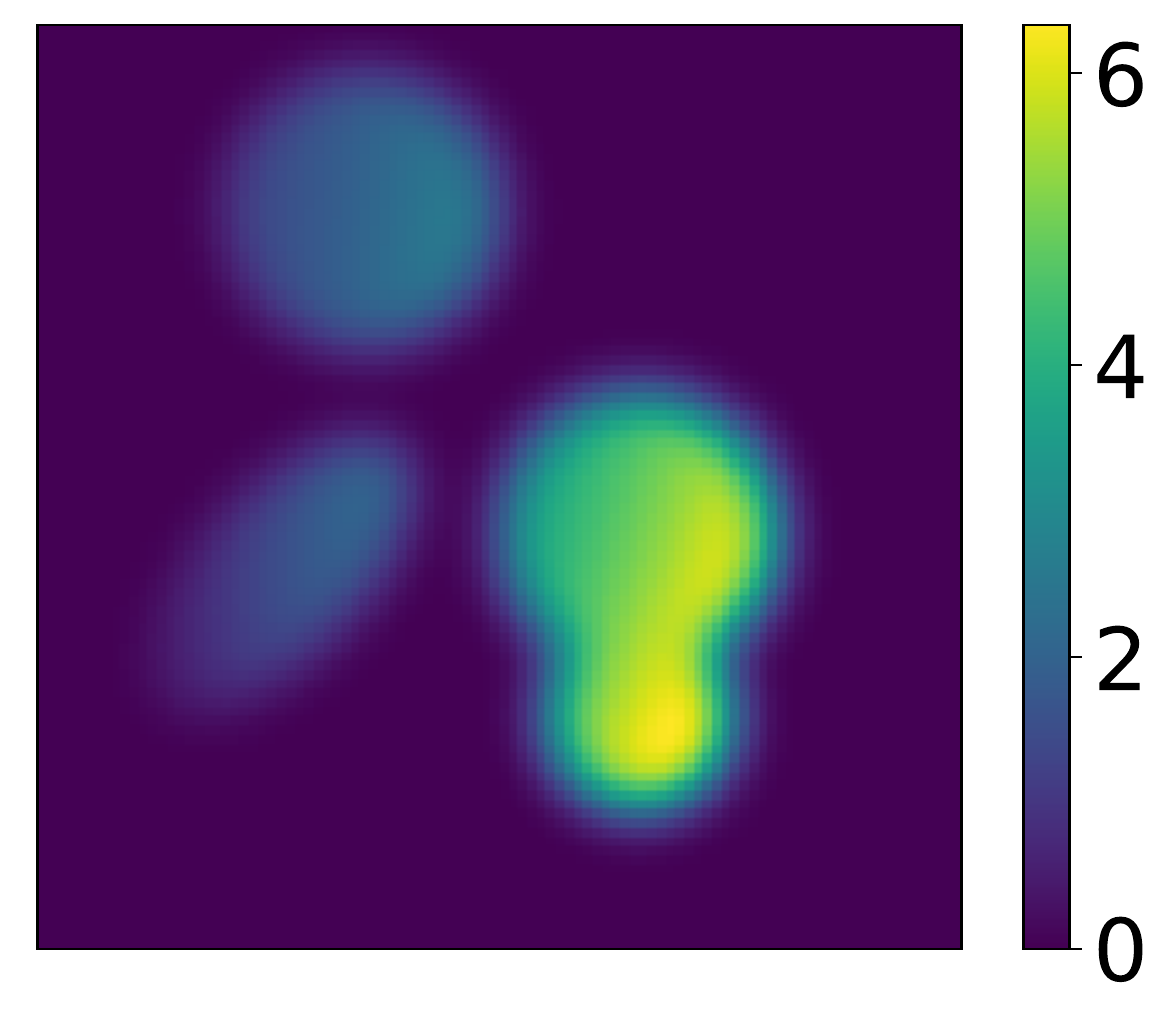} 
    \caption{Deconvolution of the function.}   
\end{subfigure}  
\includegraphics[width=0.6\textwidth]{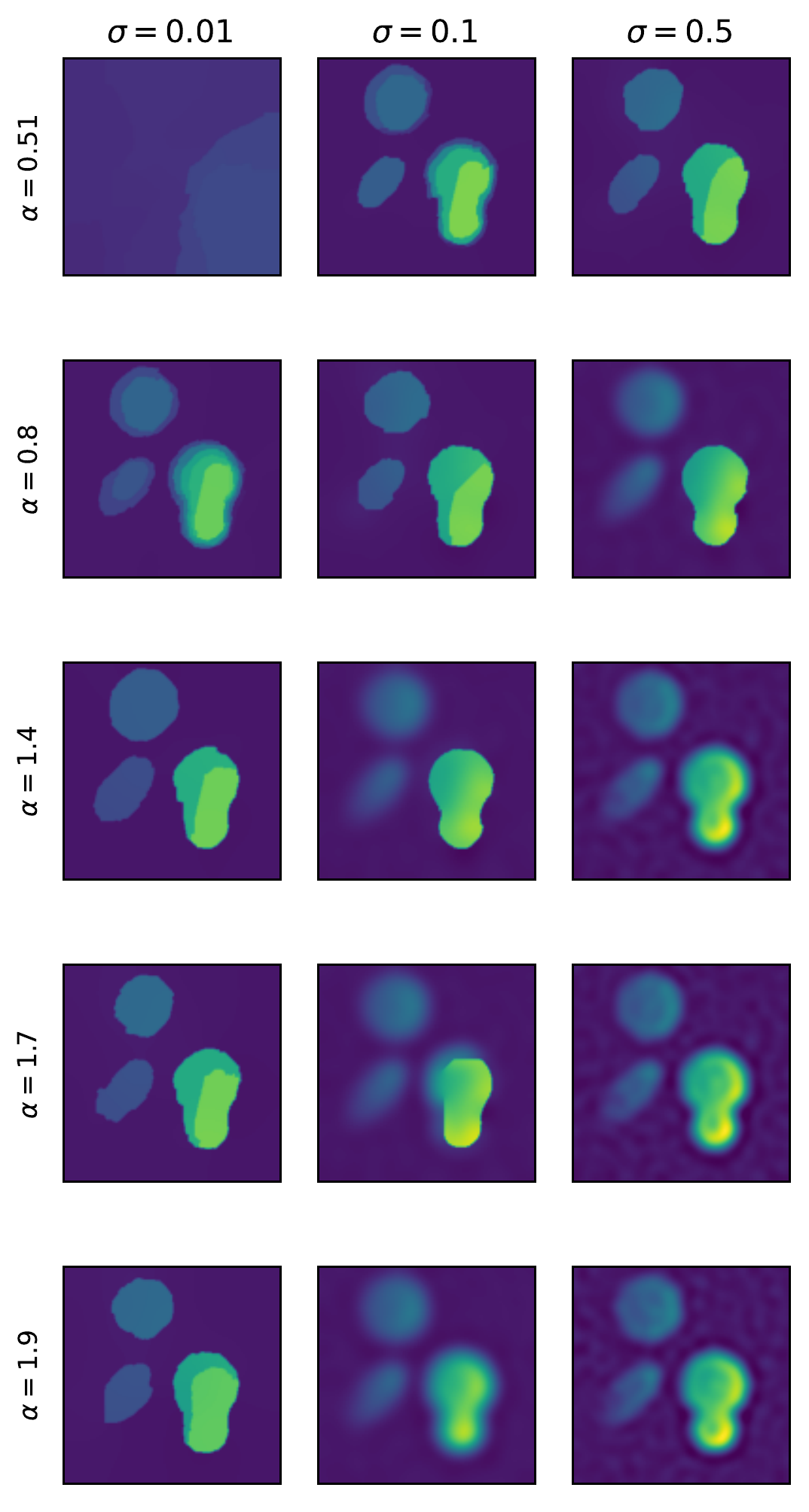}
\caption{Ground truth function, its deconvolution, and the MAP estimate reconstructions of the two-dimensional deconvolution problem with isotropic $\alpha$-stable difference priors with varying scale $\sigma$ and stability $\alpha$.}
\label{deconvs2}
\end{figure}

We also conduct a deconvolution experiment in two dimensions with similar settings as in the one-dimensional case. The  ground truth function and the reconstructions are plotted  in   Figure \ref{deconvs2}. We aim to reconstruct the blurred test function with the help of spherically symmetric bivariate $\alpha$-stable first-order difference priors \eref{bivariate}. The ground truth function is supported on $[-1,1]^2$. It is evaluated at a uniform equispaced grid of size $333\times 333$, after which is interpolated at $100^2$ points that are scattered according to a low-discrepancy sequence within the domain of the target function. 
The dataset $\mathbf{y}$ is then generated with the help of a matrix approximation for the convolution with a Gaussian kernel of $\psi$:
\[
\psi(\mathbf{p}) = \frac{150}{\pi} \exp \left(-150 \Vert\mathbf{p}\Vert^2 \right),
\]
and noncorrelated Gaussian noise with variance of $0.05^2$ is added to the result, analogously to Equation (\ref{suoramalli}).
We use the interpolation and the non-gridded evaluation points to ensure that there are no artifacts in the reconstructions that could be explained by too-regular lattice discretizations of $\mathbf{u}$ in the data generation step. 
An equispaced grid of $256\times 256$ nodes is used in the MAP estimators to prevent committing an inverse crime, and we employ a matrix approximation for the convolution model also in the reconstruction step.

The MAP estimates of the reconstructions are consistent with the one-dimensional deconvolution experiment. Increasing the stability of the $\alpha$-stable difference priors manifests in more  Gaussian-like features in the MAP estimates. The distribution   $\alpha=0.51$ and $\sigma=0.01$ is probably too spiky and heavy-tailed as the difference prior since the reconstruction lacks any features resembling the ground truth objects.  A notable feature is the existence of diagonal discontinuities at certain MAP estimates, like in the case $\alpha=0.8$ and $\sigma=0.1$ in the object that consists of two overlapping spheres. Although the construction of the $\alpha$-stable difference prior incorporates bivariate symmetrically contoured $\alpha$-stable distributions, the prior is likely not fully isotropic. In fact, even the isotropic and upwind total variation priors are not perfectly isotropic, and a method has been proposed to alleviate the issue \cite{C17}.  Unfortunately, the technique cannot be applied to the presented $\alpha$-stable priors, so the matter of alleviating the diagonally anisotropic reconstructions must be considered separately.

%

\newcommand{\qqhei}{3.4cm}
\newcommand{\pphei}{3.4cm}
\newcommand{\bboxi}{1pt}
\newcommand{\nostoa}{3pt}
\newcommand{\bboxip}{5pt}
\newcommand{\nostoap}{1.6cm}
\newcommand{\nostoaq}{0.1cm}
\newcommand{\nostoat}{0.1cm}

\subsection{Inversion of an elliptic partial differential equation}
As the third numerical experiment with the $\alpha$-stable priors, we consider the nonlinear inverse problem of estimating a conductivity field $k \in L^{\infty}(\Omega)$, with Lipschitz domain $\Omega \subset \mathbb{R}^2$ of an elliptic partial differential equation (PDE):
\begin{align}
\label{pde}   
    -\nabla \cdot (k\nabla u&) = g,  \quad x \in \Omega \\
    &u = 0,  \quad x \in \partial \Omega,  
\end{align}
with prescribed zero Dirichlet boundary conditions, where $u \in H^1_0(\Omega)$ denotes the solution of the PDE, and $g\in L^{\infty}(\Omega)$. The inversion is done using noisy observations $\mathbf{y}$ of the solution of the PDE as the likelihood for $k$. We discretize the PDE with the standard finite difference method. The noise is assumed to be Gaussian with observations taking the form 
\begin{equation}\label{noisepde}
y_i = u_i + \epsilon_i, \quad \epsilon_i \sim \mathcal{N}(0,0.001^2).
\end{equation}
Like in the other two experiments, we  evaluate only the MAP estimator of the problem. We use a bounded limited-memory BGFS algorithm (L-BFGS-B) to calculate the MAP estimate. We employ the constraint  $ 10^{-5} < k < 10^2$ to ensure the well-posedness of the elliptic PDE and to keep the condition number of the matrix of the system of equations small.  The gradient of the log posterior with respect to the discretized $k$ is calculated through a discrete adjoint method \cite{D09, G21}. In other words, we solve the adjoint equation to obtain the adjoint $\mathbf{q}$  via the equation
\begin{equation}\label{adjoint}
\left( \frac{\partial \mathbf{E}}{\partial \mathbf{u}}\right)^T \mathbf{q} = \left(\frac{\partial \pi\left(\mathbf{y}\mid\mathbf{u}(\mathbf{k})\right)}{\partial \mathbf{u}}\right)^T,
\end{equation}
where $\mathbf{E}$ denotes the system of finite difference equations of the discretized PDE, and $\pi\left(\mathbf{y}\mid\mathbf{u}(\mathbf{k})\right)$ the Gaussian likelihood function, which merely consists of solving the PDE with given $\mathbf{k}$ and evaluating the fidelity of the obtained solution with respect to $\mathbf{y}$. The gradient of the log posterior $\pi(\mathbf{k}\mid\mathbf{y})$ with respect to the discretized conductivity field $\mathbf{k}$ is then
\begin{equation}\label{gradient}
\frac{\partial \pi(\mathbf{k}\mid\mathbf{y})}{\partial \mathbf{k}} = -\mathbf{q} \frac{\partial \mathbf{E}}{\partial \mathbf{k} },
\end{equation}
since the likelihood depends on $\mathbf{k}$ only through $\mathbf{u}$. 

\begin{figure}
\centering
\begin{subfigure}[t]{5.4cm}
    \centering
    \includegraphics[width=3.4cm]{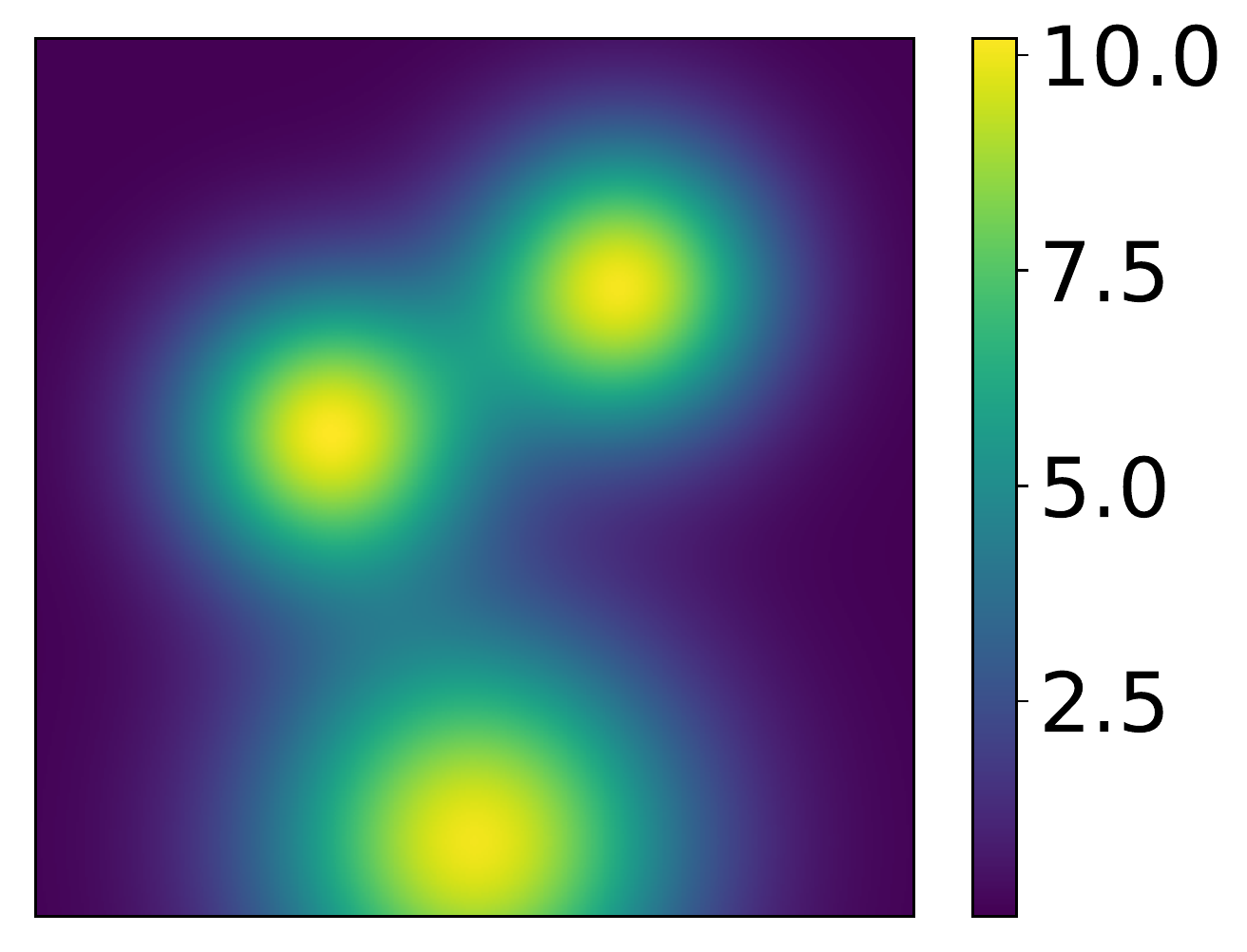}  
    \caption{Source term function $g$.}    
\end{subfigure}\begin{subfigure}[t]{5.4cm}
    \centering
    \includegraphics[width=3.4cm]{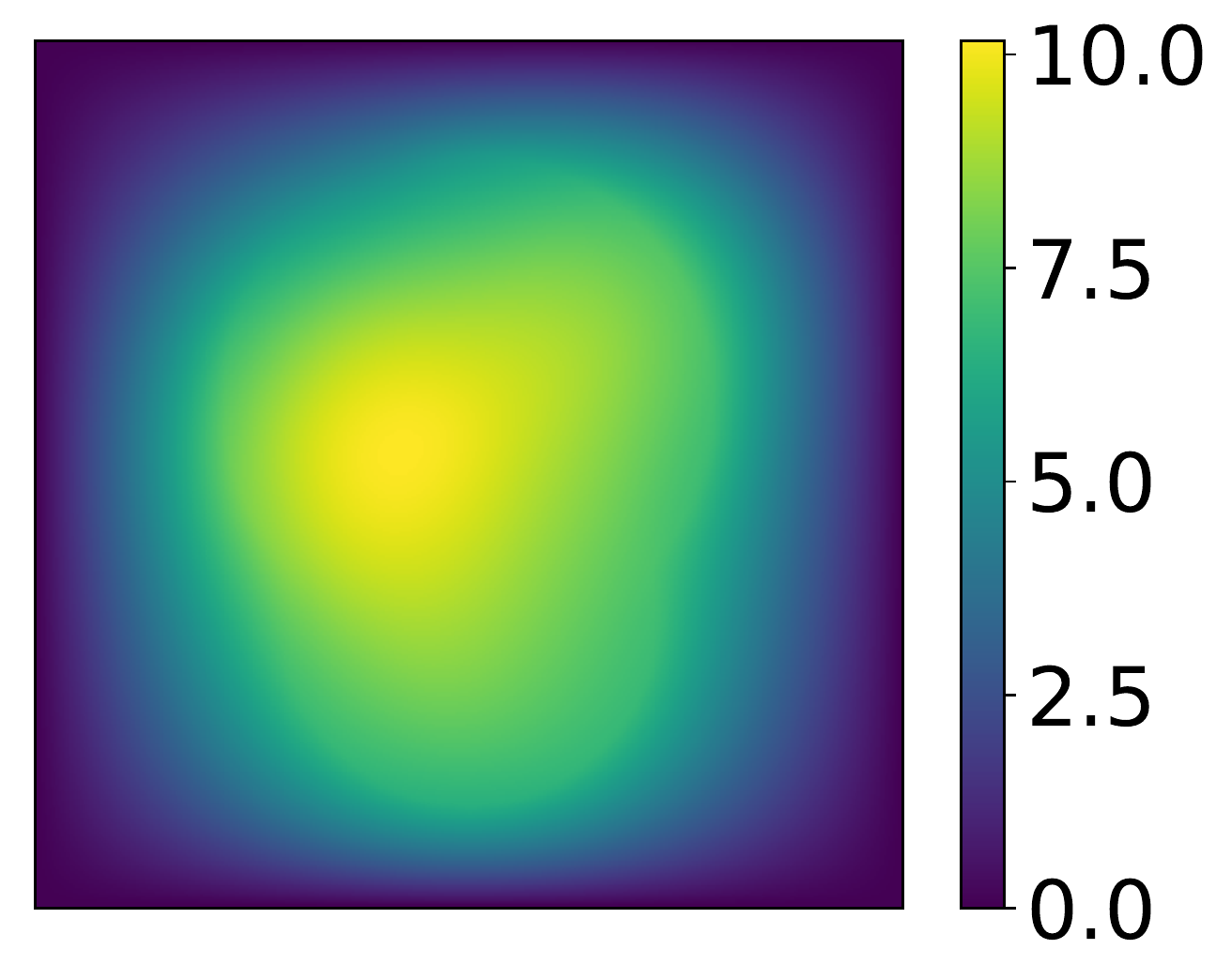}    
    \caption{Solution of the PDE with the true $k$.}  
\end{subfigure}\begin{subfigure}[t]{5.4cm}
    \centering
    \includegraphics[width=3.25cm]{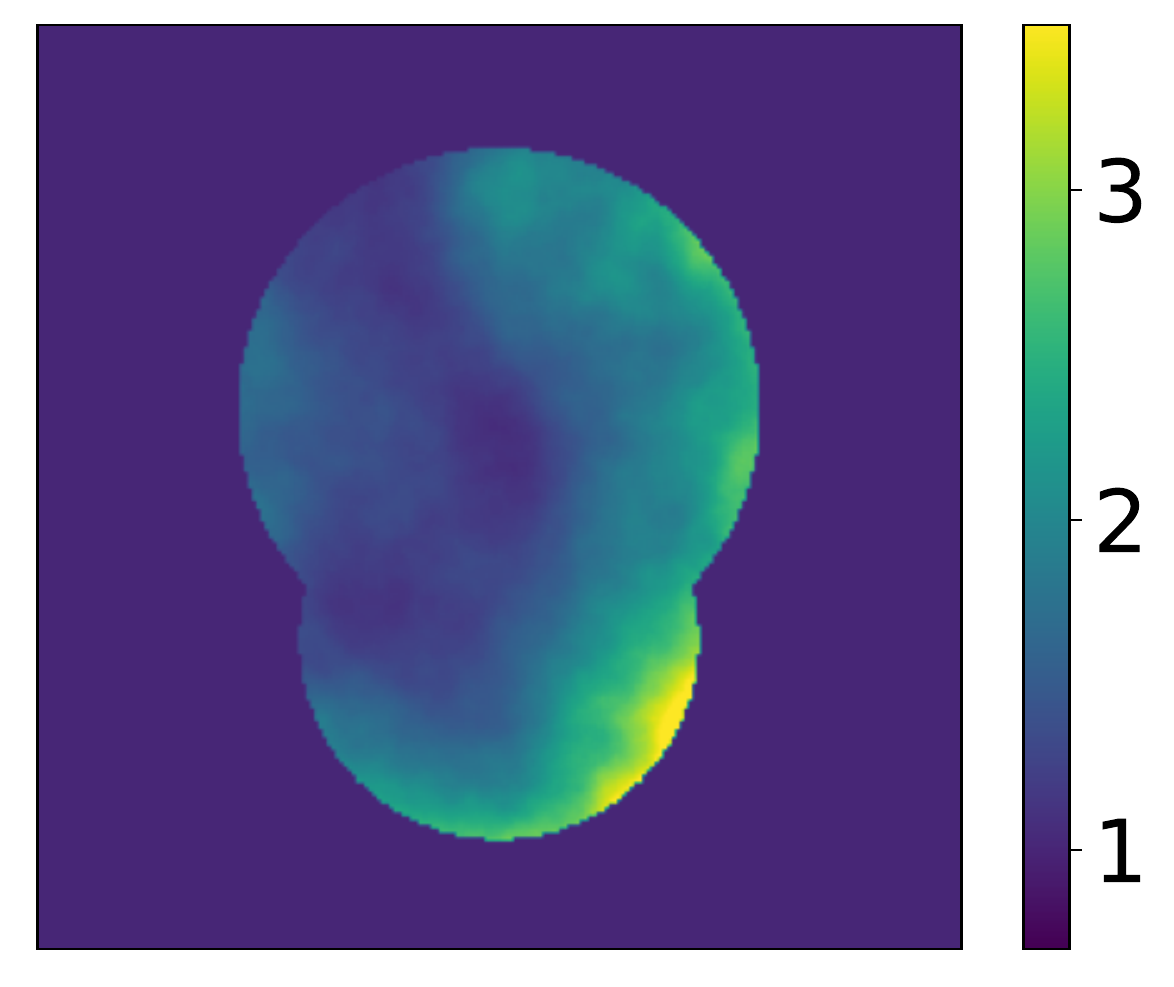} 
    \caption{True conductivity $k$.}   
\end{subfigure}

\includegraphics[width=0.6\textwidth]{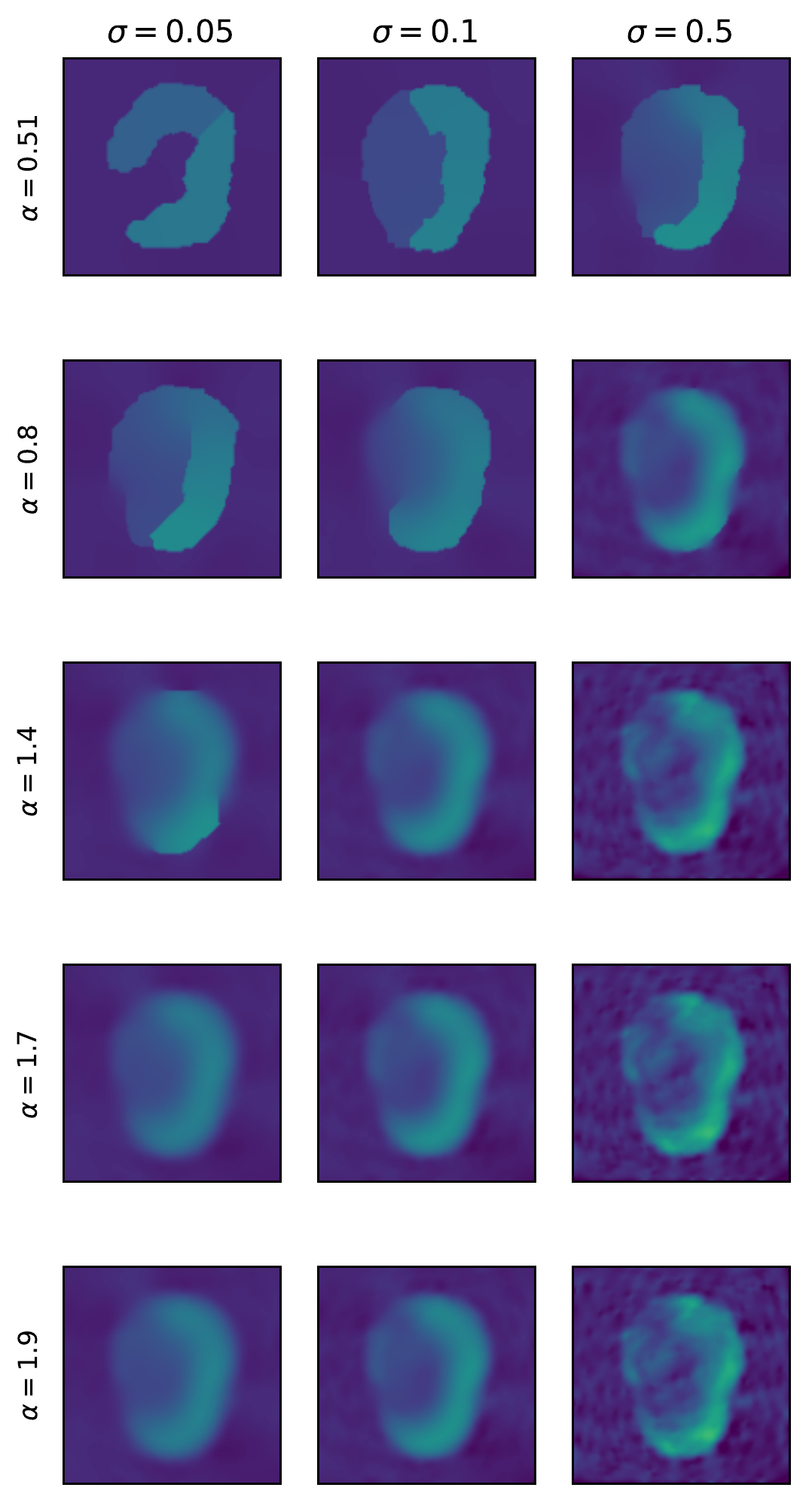}
\caption{Functions used in the nonlinear inversion of the conductivity function $k$ of an elliptic PDE, and the MAP  estimates with symmetric $\alpha$-stable difference priors with varying choices of the scale $\sigma$ and stability $\alpha$.}
\label{pderesults}
\end{figure}

We estimate the conductivity field with the same bivariate $\alpha$-stable difference priors as we do in the two-dimensional deconvolution. We use a reconstruction grid of $128\times128$. To simulate the measurements, and to avoid committing an inverse crime, we calculate the solution of the PDE using a larger finite difference grid with a size of $223\times223$ and interpolate the solution at $25\times25$ points that are positioned on the reconstruction grid according to a low-discrepancy sequence, after we add the noise to the dataset according to Equation~\eref{noisepde}. 
The source term function $g$ of  \eref{pde}, the solution of the PDE, and the ground truth conductivity, as well as  the reconstructions, are plotted in Figure \ref{pderesults}.    

The shape of a double-sphere object in the conductivity field is captured the best with the priors with smaller stability indices while increasing the stability seems to favor smooth reconstructions. On the other hand, having a large scale $\sigma$ may make the prior uninformative. Judging by the shape and distribution of the values within the reconstruction, the  prior with $\alpha=0.8$ and scale $\sigma=0.1$ could be the best  of the tested parameter choices in this case. 

\section{Conclusion}\label{sec:conc}

This work was motivated by the desire to implement approximations of the $\alpha$-stable random field priors for Bayesian inverse problems.
Because both the Cauchy and Gaussian fields are special cases of the $\alpha$-stable random fields, our objective was to extend the prior selection to general $\alpha$-stable priors, which could prove useful in reconstructions where both Gaussian and non-Gaussian features are present. 
As the $\alpha$-stable density functions mostly lack  the closed-form expressions, we  introduced a computationally feasible hybrid method for approximating the symmetric univariate and bivariate $\alpha$-stable probability density functions. 
The novelty of the presented method in comparison with the existing approximation methods is its accuracy and, especially, its performance. 
The method allows evaluation of the $\alpha$-stable probability log-density functions within a stability index range of $\alpha\in[0.5,1.9]$ and radius argument range of $r \in [0,\infty)$. 
Furthermore, we provided error bounds for the log-density approximations through careful computer-assisted analysis. The range of the stability index and the relative accuracy of the hybrid method could be potentially improved by introducing more spline grids for large $\alpha$ and small $\alpha$. Another further improvement would be to extend the domain of the splines to higher $r$ and optimize the spacing of the spline grids for greater accuracy.  

In the numerical experiments, we employed finite-difference approximations of the $\alpha$-stable first-order random motion priors at one-dimensional and two-dimensional deconvolution, and we also addressed the estimation of a function governed by an elliptic partial differential equation with the same priors. The MAP estimation was implemented through the standard L-BFGS method and its bounded variant. 
Our objective was to illustrate how the parameters, such as the stability and scale of the $\alpha$-stable priors, can affect the estimation of the unknown functions. 
The results are promising in the sense that the presented priors are both computationally viable, manifest in useful MAP estimates, and are yet novel compared to the existing random field priors like Gaussian, Cauchy, Besov, and total variation priors. 
We also introduced hierarchical $\alpha$-stable priors in our one-dimensional deconvolution example. 
Although the results are already encouraging, we believe the full potential of the hierarchical $\alpha$-stable priors is yet to be found. An approximation method like ours is needed for hierarchical $\alpha$-stable priors because the density functions must be evaluated on a continuous range of stability indices. 


As we introduced new $\alpha$-stable  priors and provided examples through MAP-estimates, we consider extending the estimators to full inference as well as other $\alpha$-stable priors. 
For future work, we will consider Bayesian neural networks \cite{RMN96} with $\alpha$-stable weights, which are possibly non-symmetric in contrast to all the distributions in this study. 
We believe the developed approximations to turn out useful in that case due to the recent studies on Bayesian neural networks with Cauchy and Gaussian weights \cite{LDS22}. 
Alternatively, $\alpha$-stable random field approximations through  the stochastic partial differential equation approach could be beneficial \cite{DB14,LRL11,SCLR21}. 
Another consideration would be to test these priors on ensemble Kalman methods \cite{CIRS18,CSW19,CST19}, which have been used and tested with hierarchical Cauchy processes.

	\section*{Acknowledgments}
	We thank Simo Särkkä and Sari Lasanen for useful and interesting discussions. 
    This work has been funded by the Research Council of Finland (project number 353095).
    
    The authors declare no conflict of interest. All the authors participated in the presented work with the four criteria as required per IOP Publishing, and have read and accepted the manuscript before submission.

	\appendix
	
	\section{Error bounds}
	
	We establish quantitative pointwise bounds for symmetric stable density functions $f(r;\alpha)$ and their partial derivatives, based on series expansions due to Bergstr\"om.
	
	In this appendix, the symbol $\remainder$ stands for a generic remainder term associated to a partial series expansion of a given density function. In particular, the symbol $\remainder$ will generally have a \emph{different} meaning from one line to another. Its precise meaning will always be clear from context.
	
	\section*{Univariate bounds for $r \to 0$}
	
	Recall that for $r > 0$, the density is given by the inverse Fourier transform as follows:
	
	\begin{equation}\label{eq:univariate-inverse-fourier}
		f(r;\alpha) = \frac{1}{\pi} \int_{0}^{\infty} \cos(r t) \rme^{-t^\alpha} \rmd t = \frac{1}{\pi} \Re \Bigl\{ \int_{0}^{\infty} \rme^{-\rmi r t - t^\alpha} \rmd t \Bigr\}.
	\end{equation}
	
	Let us first recall the asymptotic expansion for $r \to 0^+$. We may apply the Taylor series expansion for the cosine function in the first integral in \eqref{eq:univariate-inverse-fourier}, yielding
	\begin{align*}
		f(r;\alpha)
		& = \frac{1}{\pi} \sum_{k=0}^n \frac{(-1)^k r^{2k}}{(2k)!} \int_{0}^{\infty} t^{2k} \rme^{-t^\alpha} \rmd t + \remainder(r;\alpha) \\
		& = \frac{1}{\pi\alpha} \sum_{k=0}^n \frac{(-1)^k \Gamma(\frac{2k+1}{\alpha})}{(2k)!} r^{2k} + \remainder(r;\alpha),
	\end{align*}
	for any $n \in \nanu_0$. Since $\|\cos^{(\ell)}\|_{L^\infty(0,\infty)} \leq 1$ for all $\ell \in \nanu$, we can apply Taylor's theorem and arrive at the following estimate for the remainder term:
	\[
	|\remainder(r;\alpha)| \leq \frac{r^{2n+2}}{\pi (2n+2)!} \int_{0}^{\infty} t^{2n+2} \rme^{-t^\alpha} \rmd t = \frac{\Gamma(\frac{2n+3}{\alpha})}{\pi\alpha (2n+2)!} r^{2n+2}.
	\]
	
	Similarly, for $\ell \in \nanu$ we can differentiate the first integral in \eqref{eq:univariate-inverse-fourier} with respect to $r$ to get
	\begin{equation}\label{eq:univariate-partial-r}
		\frac{\partial^\ell}{\partial r^\ell} f(r;\alpha) = \frac{(-1)^{\lceil\frac{\ell}{2}\rceil}}{\pi\alpha} \sum_{k=0}^n \frac{(-1)^k\Gamma(\frac{2k+1+2\lceil\frac{\ell}{2}\rceil}{\alpha})}{(2k+o(\ell))!} r^{2k+o(\ell)} + \remainder(r;\alpha),
	\end{equation}
	where $o(\ell) = 1$ if $\ell$ is odd and $o(\ell) = 0$ otherwise, and
	\[
	|\remainder(r;\alpha)| \leq \frac{\Gamma(\frac{2n+3+2\lceil\frac{\ell}{2}\rceil}{\alpha})}{\pi\alpha(2n+2+o(\ell))!} r^{2n+2+o(\ell)}.
	\]
	
	For partial derivatives with respect to $\alpha$, we can use the decomposition \eqref{eq:univariate-partial-r} for $\ell_1 \in \nanu_0$, $\ell_2 \in \nanu$ and $\ell \defeq \ell_1+\ell_2 \in \nanu$ to get a decomposition of the form
	\begin{equation}\label{eq:univariate-mixed}
		\frac{\partial^\ell}{\partial r^{\ell_1} \partial \alpha^{\ell_2}} f(r;\alpha)
		= \frac{(-1)^{\lceil\frac{\ell_1}{2}\rceil}}{\pi} \sum_{k=0}^n \frac{(-1)^k}{(2k+o(\ell_1))!} \frac{\partial^{\ell_2}}{\partial \alpha^{\ell_2}}\Bigl[\frac{\Gamma(\frac{2k+1+2\lceil\frac{\ell_1}{2}\rceil}{\alpha})}{\alpha}\Bigr] r^{2k+o(\ell_1)} + \remainder(r;\alpha).
	\end{equation}
	Here the partial derivatives in the summands can be computed explicitly in terms of polygamma functions, and the remainder term can be estimated as
	\begin{equation}\label{eq:univariate-mixed-r}
		|\remainder(r;\alpha)| \leq \frac{r^{2n+2+o(\ell_1)}}{\pi\alpha^{\ell_2+1}(2n+2+o(\ell_1))!} \int_{0}^{\infty} |\log(t)|^{\ell_2} t^{\frac{2n+3+2\lceil\frac{\ell_1}{2}\rceil}{\alpha}-1} |p_{\ell_2}(t)| \rme^{-t} \rmd t,
	\end{equation}
	where $p_{\ell_2}$ stands for the polynomial of degree $\ell_2$ given by
	\[
	\frac{\partial^{\ell_2}}{\partial \alpha^{\ell_2}} [\rme^{-t^\alpha}] = \log(t)^{\ell_2} p_{\ell_2}(t^\alpha) \rme^{-t^\alpha},
	\]
	(and $p_0 \equiv 1$). Generally this integral cannot be evaluated exactly, but for given values of the parameters it can be estimated numerically rather efficiently, since the integrand is neither oscillating nor overly peaked.
	
	\section*{Univariate bounds for $r\to\infty$}
	
	As in \cite{HB52}, the latter integral in \eqref{eq:univariate-inverse-fourier} can be rotated from the positive real axis to the line $\{ \tau \rme^{\rmi \varphi} \,:\, \tau > 0\}$ for an arbitrary $\varphi \in (-\frac{\pi}{\max(2\alpha,1)},0)$, resulting in
	\begin{equation}\label{eq:univariate-complex-integral}
		f(r;\alpha) = \frac{1}{\pi} \Re \Bigl\{ \rme^{\rmi \varphi}\int_0^\infty \rme^{\rme^{\rmi \beta_2} r \tau} \rme^{ \rme^{\rmi \beta_1 }\tau ^\alpha } \rmd \tau \Bigr\},
	\end{equation}
	where $\beta_1 \defeq \beta_1(\alpha) \defeq \pi + \alpha\varphi \in (\frac{\pi}{2},\pi)$ and $\beta_2 \defeq \frac{3\pi}{2} + \varphi \in (\frac{\pi}{2},\frac{3\pi}{2})$. Expanding the term $\rme^{ \rme^{\rmi \beta_1 }\tau ^\alpha}$ and performing some elementary calculations for the summands yields
	\begin{align}
		f(r;\alpha) & = \sum_{k=1}^n \frac{(-1)^{k+1}\Gamma(k\alpha+1)\sin(\frac{k\alpha \pi}{2})}{\pi k!} r^{-k\alpha - 1}  \notag
		\\ & \qquad \qquad + \underbrace{\frac{1}{\pi} \Re \Bigl\{ \rme^{\rmi (\varphi + (n+1)\beta_1) }\int_0^\infty \tau^{(n+1)\alpha} M_{n+1}(\rme^{\rmi \beta_1 }\tau ^\alpha) \rme^{\rme^{\rmi \beta_2} r \tau} \rmd \tau \Bigr\}}_{\eqdef \remainder(r;\alpha)}, \label{eq:univariate-inf-expansion-1}
	\end{align}
	for all $n \in \nanu$, where $M_{n+1}$ is (the analytic continuation of) the function $z \mapsto (e^z - \sum_{k=0}^n \frac{z^k}{k!})/z^{n+1}$. Writing $M_0(z) = \rme^{z}$ for notational convenience, the above expansion also holds for $n \in \{-1,0\}$, with the understanding that the sum is zero in this case.
	
	The functions $M_{k}$ above satisfy the bound $|M_{k}(z)| \leq \frac{1}{k!}$ for $z$ with negative real part, as can be easily verified for $k = 0$ and consequently proved inductively using the recursive formula $\frac{\rmd}{\rmd z}[z^k M_k(z)] = z^{k-1} M_{k-1}(z)$ for $k \geq 1$. Since the term $\rme^{i\beta_1} \tau^\alpha$ in the integral in \eqref{eq:univariate-inf-expansion-1} has a negative real part, the error term can thus be estimated by
	\[
	|\remainder(r;\alpha)| \leq \frac{1}{\pi(n+1)!} \int_{0}^{\infty} \tau^{(n+1)\alpha} \rme^{\sin(\varphi) r \tau} \rmd \tau = \frac{\Gamma((n+1)\alpha+1)}{\pi(n+1)! |\sin(\varphi)|^{(n+1)\alpha+1}} r^{-(n+1)\alpha-1},
	\]
	and since the true value of $\remainder$ does not actually depend on the auxiliary parameter $\varphi \in (-\frac{\pi}{\max(2\alpha,1)},0)$, the above estimate can be improved to
	\begin{equation}\label{eq:univariate-inf-remainder-1}
		|\remainder(r;\alpha)| \leq \frac{\Gamma((n+1)\alpha+1)}{\pi(n+1)! \sin(\pi_\alpha)^{(n+1)\alpha+1}} r^{-(n+1)\alpha-1},
	\end{equation}
	where $\pi_\alpha \defeq \frac{\pi}{2\max(\alpha,1)}$.
	
	Differentiating \eqref{eq:univariate-inf-expansion-1} termwise with respect to $r$, we have
	
	\begin{equation}\label{eq:univariate-inf-r}
		\frac{\partial^\ell}{\partial r^{\ell}} f(r;\alpha)
		= \frac{(-1)^{\ell}}{\pi}\sum_{k=1}^n \frac{(-1)^{k+1} }{k!} \frac{\Gamma(k\alpha+\ell+1)\sin(\frac{k\alpha \pi}{2})} {r^{k\alpha +\ell+ 1}} + \remainder(r;\alpha),
	\end{equation}
	with
	\begin{equation}\label{eq:univariate-inf-r-remainder}
		|\remainder(r;\alpha)| \leq \frac{\Gamma((n+1)\alpha+\ell+1)}{\pi(n+1)! \sin(\pi_\alpha)^{(n+1)\alpha+\ell+1}} r^{-(n+1)\alpha-\ell-1}.
	\end{equation}
	
	Finally, for pure and mixed partial derivatives with respect to $\alpha$, we may use the previous expansion as a stepping stone to obtain
	
	\begin{equation}\label{eq:univariate-inf-mixed}
		\frac{\partial^\ell}{\partial r^{\ell_1} \partial \alpha^{\ell_2}} f(r;\alpha)
		= \frac{(-1)^{\ell_1}}{\pi}\sum_{k=1}^n \frac{(-1)^{k+1} }{k!} \frac{\partial^{\ell_2}}{\partial\alpha^{\ell_2}}\Bigl[\frac{\Gamma(k\alpha+\ell_1+1)\sin(\frac{k\alpha \pi}{2})} {r^{k\alpha +\ell_1+ 1}}\Bigr] + \remainder(r;\alpha),
	\end{equation}
	with
	\begin{equation}\label{eq:univariate-inf-r-mixed}
		|\remainder(r;\alpha)| \leq \frac{1}{\pi} \int_{0}^{\infty} \tau^{\ell_1} \big|\frac{\partial^{\ell_2}}{\partial \alpha^{\ell_2}}\bigl[ \rme^{\rmi(n+1)\beta_1} \tau^{(n+1)\alpha}M_{n+1}\bigl(\rme^{\rmi \beta_1}\tau^{\alpha}\bigr)\bigr] \big| \rme^{\sin(\varphi)r\tau} \rmd \tau.
	\end{equation}
	In \eqref{eq:univariate-inf-mixed}, the derivatives in the summands can again be computed in terms of the polygamma functions if necessary. In \eqref{eq:univariate-inf-r-mixed}, the partial derivatives can be estimated in terms of the functions $M_k$ introduced above by iterating the recursive formula $M_k' = M_k - k M_{k+1}$. The integral in \eqref{eq:univariate-inf-r-mixed} will be of the order $\mathcal{O}(r^{-(n+1)\alpha-\ell_1-1} \log(r)^{\ell_2})$ for large values of $r$, and one can a posteriori take $\varphi \to -\pi_\alpha$ since again the true value of $\remainder(r;\alpha)$ does not depend on $\varphi$.
	
	\section*{Bivariate bounds for $r \to 0$}
	
	A random two-dimensional vector $\bi{X}$ obeying a symmetric and \emph{spherically contoured} bivariate stable distribution with stability index $\alpha \in (0,2)$ can be described in terms of its characteristic function:
	\[
	\E[\exp(\rmi \bi{t}^{\intercal} \bi{X})] = \exp(-|\bi{X}|^\alpha), \qquad \bi{t} \in \real^2,
	\]
	where $|\cdot|$ refers to the standard $\ell^2$-based Euclidean norm. We refer to e.g.~\cite{JPN13} for a comprehensive account on such distributions.
	
	The density function $f_{\bi{X}}$ of $\bi{X}$ can be expressed at $\bi{x} \in \real^2$ by the 
	inverse Fourier transform of the characteristic function above, resulting in
	\[
	f_{\bi{X}}(\bi{x};\alpha) = \frac{1}{2\pi} \int_{0}^{\infty} J_0(|\bi{x}| t) t \rme^{-t^\alpha} \rmd t;
	\]
	see e.g.~\cite{KZ04}. Here $J_\nu$ with $\nu = 0$ stands for the Bessel function of the first kind, which can for $\nu \in \nanu_0$ and $z \in \complex$ be expressed as
	\begin{equation}\label{eq:bessel-nu}
		J_\nu(z) = \sum_{k=0}^{\infty} \frac{(-1)^k}{k!(k+\nu)!}\Bigl(\frac{z}{2}\Bigr)^{2k+\nu}.
	\end{equation}
	
	In this section we are interested in estimating partial derivatives of the density function $f_{\bi{X}}$ in terms of $|\bi{x}|$ and $\alpha$. For this purpose, we write
	\begin{equation}\label{eq:bivariate-besselj0}
		f(r;\alpha) \defeq \frac{1}{2\pi}\int_{0}^{\infty} J_0(rt) t \rme^{-t^\alpha} \rmd t, \quad r \geq 0,
	\end{equation}
	for the radial density function of $\bi{X}$.
	
	By well-known integral representation formulas \cite[Equation (4.9.11)]{AAR}, we have the uniform bounds $\|J_\nu\|_{L^\infty(\real)} \leq 1$ on the real line for all $\nu \in \nanu_0$, and by standard recurrence relations \cite[Equation (4.6.6)]{AAR} we thus have $\| J^{(\ell)}_\nu \|_{L^{\infty}(\real)} \leq 1$ for all $\nu \in \nanu_0$ and derivatives $J^{(\ell)}_\nu$ of $J_\nu$. Hence we may expand the function $J_0$ in \eqref{eq:bivariate-besselj0} to obtain
	\[
	f(r;\alpha) = \frac{1}{2\pi\alpha} \sum_{k=0}^{n} \frac{(-1)^k\Gamma(\frac{2k+2}{\alpha})}{(2^k k!)^2} r^{2k} + \remainder(r;\alpha),
	\]
	with
	\[
	|\remainder(r;\alpha)| \leq \frac{\Gamma(\frac{2n+4}{\alpha})}{2\pi\alpha(2n+2)!}r^{2n+2}.
	\]
	
	Similarly, differentiating \eqref{eq:bivariate-besselj0} with respect to $r$, and \eqref{eq:bessel-nu} for $\nu = 0$ with respect to $z$, yields
	\begin{align*}
		& \qquad \frac{\partial^\ell}{\partial r^\ell} f(r;\alpha) \\
		& = \frac{(-1)^{\lceil\frac{\ell}{2}\rceil}}{2\pi\alpha} \sum_{k=0}^n \frac{(-1)^k \bigl\{ \prod_{i=1}^{\ell} (2k+o(\ell)+i) \bigr\}\Gamma(\frac{2k+2+2\lceil\frac{\ell}{2}\rceil}{\alpha})}{(2^{k+\lceil\frac{\ell}{2}\rceil}(k+\lceil\frac{\ell}{2}\rceil)!)^2} r^{2k+o(\ell)} + \remainder(r;\alpha),
	\end{align*}
	for all $\ell \in \nanu$, where
	\[
	|\remainder(r;\alpha)| \leq \frac{\Gamma(\frac{2n+4+2\lceil\frac{\ell}{2}\rceil}{\alpha})}{2\pi\alpha (2n+2+o(\ell))!} r^{2n+2+o(\ell)}.
	\]
	Subsequently
	\begin{align*}
		\frac{\partial^\ell}{\partial r^{\ell_1} \partial \alpha^{\ell_2}} f(r;\alpha) & = \frac{(-1)^{\lceil\frac{\ell_1}{2}\rceil}}{2\pi} \sum_{k=0}^n \frac{(-1)^k \bigl\{ \prod_{i=1}^{\ell_1} (2k+o(\ell_1)+i) \bigr\}}{(2^{k+\lceil\frac{\ell_1}{2}\rceil}(k+\lceil\frac{\ell_1}{2}\rceil)!)^2} \frac{\partial^{\ell_2}}{\partial \alpha^{\ell_2}}\Bigl[\frac{\Gamma(\frac{2k+2+2\lceil\frac{\ell_1}{2}\rceil}{\alpha})}{\alpha} \Bigr]r^{2k+o(\ell_1)} \\
		& \qquad \qquad + \remainder(r;\alpha),
	\end{align*}
	for all $\ell_1 \in \nanu_0$ and $\ell_2 \in \nanu$, where
	\[
	|\remainder(r;\alpha)| \leq \frac{r^{2n+2+o(\ell_1)}}{2\pi\alpha^{\ell_2+1}(2n+2+o(\ell_1))!} \int_{0}^{\infty} |\log(t)|^{\ell_2} t^{\frac{2n+4+2\lceil\frac{\ell_1}{2}\rceil}{\alpha}-1} |p_{\ell_2}(t)| \rme^{-t}\rmd t.
	\]
	Here $p_{\ell_2}$ stands for the polynomials introduced in \eqref{eq:univariate-mixed-r} above. Again, the integrals appearing on the right-hand side well-behaved enough for numerical approximation; see the discussion following \eqref{eq:univariate-mixed-r}.
	
	\section*{Bivariate bounds for $r \to \infty$}
	
	Nolan \cite{JPN13} records an asymptotic expansion for the density function of the \emph{amplitude distribution} of $\bi{X}$ as $r \to \infty$, which for the radial density function \eqref{eq:bivariate-besselj0} translates as
	\begin{equation}\label{eq:nolan-asymptotic}
		f(r;\alpha) = \frac{1}{\pi^2} \sum_{k=1}^{n} \frac{(-1)^{k+1}2^{k\alpha}\Gamma(\frac{k\alpha+2}{2})^2 \sin(\frac{k\alpha\pi}{2})}{k!}r^{-k\alpha-2} + \mathcal{O}\bigl( r^{-(n+1)\alpha-2}\bigr),
	\end{equation}
	for $\alpha \in (0,2)$, $r > 0$ and $n \in \nanu$. This can be obtained by expressing $\bi{X}$ as a sub-Gaussian vector with respect to a certain totally skewed univariate stable distribution, which admits a similar asymptotic series expansion as described in \cite{HB52}. Below we will obtain \eqref{eq:nolan-asymptotic} in an alternative way that allows us to quantitatively control the error term.
	
	We may rewrite \eqref{eq:bivariate-besselj0} as
	\begin{equation}\label{eq:bivariate-hankel0}
		f(r;\alpha) = \frac{1}{2\pi} \Re\Bigl\{ \int_{0}^{\infty} H_0(r t) t \rme^{-t^\alpha} \rmd t\Bigr\},
	\end{equation}
	where $H_0 \colon \complex \setminus (-\infty,0] \to \complex$ stands for the so-called Hankel function of the first kind \cite[Section 4.7]{AAR}, defined in terms of the Bessel functions of the first kind ($J_0$) and second kind ($Y_0$) by
	\[
	H_0(z) = J_0(z) + \rmi Y_0 (z).
	\]
	The functions $H_\nu$, $\nu \in \nanu_0$, can be defined similarly in terms of $J_\nu$ and $Y_\nu$. By well-known recurrection relations and connection formulas \cite[Sections 10.6 and 10.4]{DLMF}, each derivative $H_{\nu}^{(\ell)}$ can again be expressed as a linear combination of functions $H_{\nu'}$ with $(\nu-\ell)_+ \leq \nu' \leq \nu+\ell$.
	
	Following the contour integration procedure described in \cite{HB52}, we can then rotate the integral in \eqref{eq:bivariate-hankel0} from the positive real axis to the line $\{\rme^{i\varphi} \tau \,:\, \tau > 0\}$ for an arbitrary $\varphi \in (0, \frac{\pi}{\max(2\alpha),1})$ to get
	\begin{equation}\label{eq:bivariate-complex-integral}
		f(r;\alpha) = \frac{1}{2\pi} \Re\Bigl\{ \rme^{2\rmi \varphi} \int_{0}^{\infty} H_0\bigl(\rme^{\rmi\varphi} r\tau \bigr) \tau \rme^{\rme^{\rmi \beta_1} \tau^\alpha} \rmd \tau\Bigr\},
	\end{equation}
	where $\beta_1 \defeq \beta_1(\alpha) \defeq \pi + \alpha\varphi$ is as in \eqref{eq:univariate-complex-integral}. More precisely, this contour integration and the associated limiting procedure is permitted because
	\begin{equation}\label{eq:hankel-bounds}
		|H_{\nu}(z)| \leq
		\begin{cases}
			c_{1,\nu} \bigl(|z|^{-\nu} + |\log(z)|\bigr) & \quad \text{for} \quad 0 < |z| \leq 1\\
			c_{2,\nu} |z|^{-\frac{1}{2}}\rme^{-\Im(z)} & \quad \text{for} \quad |z| \geq 1
		\end{cases}
	\end{equation}
	for all $\nu \in \nanu_0$, with certain constants $c_{i,\nu}$ that we will not elaborate on; we refer to \cite[Sections 4.5 and 4.8]{AAR} for this and more comprehensive asymptotic expansions for Hankel functions.
	
	Because of \eqref{eq:hankel-bounds}, we calso expand the term $\rme^{\rme^{\rmi \beta_1} \tau^\alpha}$ in \eqref{eq:bivariate-complex-integral} and integrate termwise to get
	\[
	f(r;\alpha) = \frac{1}{2\pi} \sum_{k=0}^{n} \frac{1}{k!} \Re\Bigl\{ \rme^{\rmi (2\varphi + k\beta_1)} \int_{0}^{\infty} H_0\bigl(\rme^{\rmi\varphi} \tau \bigr) \tau^{k \alpha + 1} \rmd \tau\Bigr\} r^{-k\alpha-2} + \remainder(r;\alpha).
	\]
	If we can show that the remainder term above is of the order $\mathcal{O}(r^{-(n+1)\alpha-2})$ as $r \to \infty$, it follows automatically from the uniqueness property of asymptotic expansions that the principal term coincides with the one in \eqref{eq:nolan-asymptotic}. To this end we recall the functions $M_k$ from \eqref{eq:univariate-inf-expansion-1} and write
	\[
	\remainder(r;\alpha) = \frac{1}{2\pi}\Re\Bigl\{ \rme^{\rmi (2\varphi + (n+1)\beta_1)} \int_{0}^{\infty} H_0\bigl(\rme^{\rmi\varphi} r\tau \bigr) \tau^{(n+1)\alpha + 1} M_{n+1}\bigl(\rme^{\rmi \beta_1} \tau^\alpha\bigr) \rmd \tau\Bigr\},
	\]
	and since the term $\rme^{\rmi \beta_1} \tau^\alpha$ above has a negative real part by assumption, we may estimate
	\[
	|\remainder(r;\alpha)| \leq \frac{1}{2\pi (n+1)!} \Bigl(\int_{0}^{\infty} |H_0(e^{\rmi\varphi}\tau)| \tau^{(n+1) \alpha+1} \rmd \tau \Bigr) r^{-(n+1)\alpha-2}
	\]
	which is of the desired form. In light of \eqref{eq:hankel-bounds}, we may further take $\varphi \to \frac{\pi}{2 \max(\alpha,1)} = \pi_\alpha$ so that
	\begin{equation}\label{eq:bivariate-inf-r}
		|\remainder(r;\alpha)| \leq \frac{1}{2\pi(n+1)!} \Bigl(\int_{0}^{\infty} |H_0(e^{\rmi\pi_\alpha}\tau)| \tau^{(n+1)\alpha+1} \rmd \tau \Bigr) r^{-(n+1)\alpha-2}.
	\end{equation}
	
	Similarly,
	\begin{align*}
		\frac{\partial^\ell}{\partial r^\ell} f(r;\alpha)
		& = \frac{(-1)^{\ell}}{\pi^2} \sum_{k=1}^{n} \frac{(-1)^{k+1}2^{k\alpha}\bigl\{\prod_{i=2}^{\ell+1}(k\alpha+i)\bigr\}\Gamma(\frac{k\alpha+2}{2})^2 \sin(\frac{k\alpha\pi}{2})}{k!}r^{-k\alpha-\ell-2} \\
		& \qquad \qquad + \remainder(r;\alpha)
	\end{align*}
	with
	\[
	|\remainder(r;\alpha)| \leq \frac{1}{2\pi(n+1)!} \Bigl(\int_{0}^{\infty} |H_0^{(\ell)}(e^{\rmi\pi_\alpha}\tau)| \tau^{(n+1)\alpha+\ell+1} \rmd \tau \Bigr) r^{-(n+1)\alpha-\ell-2},
	\]
	and further
	\begin{align*}
		&\frac{\partial^\ell}{\partial r^{\ell_1} \partial \alpha^{\ell_2}} f(r;\alpha)
		= \\ &\frac{(-1)^{\ell_1}}{\pi^2} \sum_{k=1}^{n} \frac{(-1)^{k+1}}{k!}\frac{\partial^{\ell_2}}{\partial \alpha^{\ell_2}}\Bigl[\frac{2^{k\alpha}\bigl\{\prod_{i=2}^{\ell+1}(k\alpha+i)\bigr\}\Gamma(\frac{k\alpha+2}{2})^2 \sin(\frac{k\alpha\pi}{2})}{r^{k\alpha+\ell_1+2}}\Bigr] 
		+ \remainder(r;\alpha),
	\end{align*}
	with
	\begin{equation}\label{eq:bivariate-inf-mixed-r}
		|\remainder(r;\alpha)| \leq \frac{1}{2\pi} \int_{0}^{\infty} \big|H_0^{(\ell_1)}\bigl(\rme^{\rmi\varphi} r\tau \bigr) \big| \tau^{\ell_1 + 1}  \Big| \frac{\partial^{\ell_2}}{\partial \alpha^{\ell_2}}\bigl[\rme^{\rmi(n+1)\beta_1}\tau^{(n+1)\alpha} M_{n+1}\bigl(\rme^{\rmi \beta_1} \tau^\alpha\bigr) \bigr] \Big| \rmd \tau.
	\end{equation}
	The latter integrand can again be estimated in terms of the functions $M_k$; see \eqref{eq:univariate-inf-r-mixed} and the relevant discussion. One will then end up with integrals that are analytically untractable, but not outside the reach of numerical estimation.
	
	\section*{Uniform and oscillatory bounds for $0 \ll r \ll \infty$}
	
	Here we present rather crude uniform bounds, and some refinements based on oscillatory integral techniques, of $|\frac{\partial^
		{\ell_1 + \ell_2}}{\partial r^{\ell_1} \partial \alpha^{\ell_2}} f(r;\alpha)|$ for ``moderate'' values of $r$, where none of the asymptotic expansions discussed earlier come close to approximating the true function with e.g.~$2$--$3$ summands. The precise range of these ``moderate'' values of $r$ of course depends on $\alpha$ and the $\ell_i$'s. We present these estimates only in the univariate case, since the bivariate case can be handled with minor adjustments.
	
	First, from \eqref{eq:univariate-inverse-fourier}, we have
	\begin{equation}\label{eq:oscillatory}
		\frac{\partial^
			{\ell_1 + \ell_2}}{\partial r^{\ell_1} \partial \alpha^{\ell_2}} f(r;\alpha) = \frac{1}{\pi} \int_{0}^{\infty} \cos^{(\ell_1)} (r t) \log(t)^{\ell_2} t^{\ell_1} p_{\ell_2}(t^\alpha) \rme^{-t^\alpha}  \rmd t,
	\end{equation}
	where $\cos^{(\ell_1)}$ stands for the $\ell_1$'th derivative of the cosine function, and $p_{\ell_2}$ stands for the polynomial introduced in \eqref{eq:univariate-mixed-r}. Thus, a simple application of the triangle inequality yields
	\[
	\Bigl| \frac{\partial^
		{\ell_1 + \ell_2}}{\partial r^{\ell_1} \partial \alpha^{\ell_2}} f(r;\alpha) \Bigr|
	\leq \frac{1}{\pi \alpha^{\ell_2+1}} \int_{0}^{\infty} |\log(t)|^{\ell_2} t^{\frac{\ell_1+1}{\alpha} - 1} |p_{\ell_2}(t)| \rme^{-t} \rmd t.
	\]
	The latter integral is again usually intractable (although it can be expressed in terms of the standard gamma function if $\ell_2 = 0$), but numerically estimable.
	
	In case $\ell_2 = 0$ and $\ell_1 \defeq \ell > 0$, we can slightly refine \eqref{eq:oscillatory} using the oscillatory integral technique of partial integration against a function that decays sufficiently fast as $t \to 0^+$ and $t \to \infty$:
	\begin{equation} \label{eq:oscillatory-r}
		\Bigl| \frac{\partial^{\ell}}{\partial r^{\ell} } f(r;\alpha) \Bigr|  = \frac{1}{\pi r} \Bigl| \int_{0}^{\infty} \cos^{(\ell-1)}(rt) \frac{\rmd}{\rmd t}\bigl[ t^{\ell} \rme^{-t^\alpha }\bigr] \rmd t \Bigr| \leq \frac{1}{rt} \int_{0}^{\infty} \bigl| \frac{\rmd}{\rmd t}\bigl[ t^{\ell} \rme^{-t^\alpha }\bigr] \bigr| \rmd t.
	\end{equation}
	The derivative inside the last integral can be easily computed, and one arrives at an upper bound for the integral that can be expressed in terms of the gamma function.
	
	Similarly, when $\ell_1 = 0$ and and $\ell_2 \defeq \ell > 0$, we have
	\[
	\Bigl| \frac{\partial^{\ell}}{\partial \alpha^{\ell}} f(r;\alpha) \Bigr|
	= \frac{1}{\pi r} \Bigl| \int_{0}^{\infty} \sin(rt) \frac{\rmd^{\ell+1}}{\rmd t \, \rmd \alpha^{\ell}}\bigl[ \rme^{-t^\alpha }\bigr] \rmd t \Bigr| \leq \frac{1}{\pi r} \int_{0}^{\infty} \bigl| \frac{\rmd^{\ell+1}}{\rmd t \, \rmd \alpha^{\ell}}\bigl[ \rme^{-t^\alpha }\bigr] \bigr| \rmd t.
	\]
	Again, the derivatives inside the last integral can be computed, and after a suitable change of variables we arrive at an integral directly directly proportional to $\alpha^{-\ell}$, where the coefficient of $\alpha^{-\ell}$ can be calculated numerically.
	
	This approach could also in principle be used for mixed derivatives of $f(r;\alpha)$, and in \eqref{eq:oscillatory-r} the partial integration trick could be applied $\ell$ times instead of once. In both cases however the computations become very unwieldy, and thus we discard them.
	
	\section*{Grid approximations}
	
	Recall from the main paper (Section 3.1) that one of our primary goals is to estimate
	\begin{equation} \label{eq:goal-supremum}
		\sup_{0 < r < 30,\; 0.5 \leq \alpha \leq 1.9}\Big| \frac{\partial^{\ell_1+\ell_2}}{\partial r^{\ell_1} \partial \alpha^{\ell_2}} \log f(r;\alpha) \Big|
	\end{equation}
	for $(\ell_1,\ell_2) \in \{(4,0), (2,2), (0,4)\}$. Recalling that e.g.
	\begin{align*}
		& \bigl(\log f(r;\alpha)\bigr)^{(4,0)} = -6 \frac{f^{(1,0)}(r;\alpha)^4}{f(r;\alpha)^4} + 12 \frac{f^{(2,0)}(r;\alpha) f^{(1,0)}(r;\alpha)^2}{f(r;\alpha)^3} \\
		& \qquad\qquad - 4 \frac{f^{(3,0)}(r;\alpha) f^{(1,0)}(r;\alpha)}{f(r;\alpha)^2} - 3 \frac{f^{(2,0)}(r;\alpha)^2}{f(r;\alpha)^2} + \frac{f^{(4,0)}(r;\alpha)}{f(r;\alpha)},
	\end{align*}
	and similarly for the other relevant partial derivatives, we may estimate the suprema in \eqref{eq:goal-supremum} by applying the triangle inequality on sums of the above sort, and by estimating
	\begin{enumerate}
		\item the absolute values of the partial derivatives $f^{(\ell_1,\ell_2)}(r;\alpha)$ from above, relying on the pointwise estimates established in the previous sections, and
		\item $f(r;\alpha)$ itself from below.
	\end{enumerate}
	In this section we will give an overview of a numerical procedure to estimate these functions in a manner that is applicable for estimating \eqref{eq:goal-supremum}.
	
	We start with a fixed grid of the $(r,\alpha)$-space, with density given by a parameter $\Delta > 0$ such that 
	\[
	\frac{1.9-0.5}{\Delta} \eqdef i^*, \quad \text{and} \quad \frac{30}{\Delta} \eqdef j^* ,
	\]
	are positive integers. In our numerical simulations we have used $\Delta \defeq \frac{2}{10^3}$ in order to conserve computational resources, but other choices are possible as well.
	
	Write then
	\[
	Q_{i,j} \defeq \bigl[ (j-1)\Delta,j\Delta\bigr] \times \bigl[0.5 + (i-1)\Delta, 0.5 + i\Delta \bigr] \quad \text{for } i \in 1{:}i^* \text{ and } j \in 1{:}j^*,
	\]
	so that
	\[
	\bigcup_{i,j} Q_{i,j} = [0,30]\times[0.5,1.9],
	\]
	and the interiors of the $Q_{i,j}$'s are pairwise disjoint. The idea is to numerically estimate $f(r;\alpha)$ and its partial derivatives uniformly in these squares, of which there is a finite amount, and the uniform estimates will not be too far off from the pointwise estimates we obtained in the earlier sections if the parameter $\Delta$ is sufficiently small.
	
	To this direction, let us first discuss the estimates of the absolute values of the partial derivatives $f^{(\ell_1,\ell_2)}(r;\alpha)$ for $(r,\alpha) \in Q_{i,j}$. Many of the pointwise estimates we have discussed above -- for example \eqref{eq:univariate-partial-r}, \eqref{eq:univariate-mixed}, \eqref{eq:univariate-mixed-r}, \eqref{eq:univariate-inf-expansion-1}, \eqref{eq:univariate-inf-remainder-1}, \eqref{eq:univariate-inf-r},\eqref{eq:univariate-inf-r-remainder},  \eqref{eq:univariate-inf-mixed}, \eqref{eq:univariate-inf-r-mixed} and the bivariate versions of these estimates -- consist of terms that are monotonous with respect to both $r$ and $\alpha$ either directly, or after some simple additional upper estimates, such as applying the triangle inequality in the sums like the ones in \eqref{eq:univariate-partial-r} and \eqref{eq:univariate-inf-r}, and considering the cases $r < 1$ and $r \geq 1$ separately for terms of the form $r^{k\alpha + \ell + 1}$. 
	
	Some additional care has to be taken for some of the more complicated terms appearing in the series expansions for the partial derivatives involving the variable $\alpha$, as well as the respective remainder terms. We explain this with examples pertaining to the univariate case, but everything here applies to the bivariate case as well with obvious modifications.
	
	First, the derivatives with respect to $\alpha$ appearing in the summands in \eqref{eq:univariate-mixed} and \eqref{eq:univariate-inf-mixed} can with some effort be computed for $\ell_2 \in \{1,2\}$, resulting terms involving the gamma function itself and the so-called polygamma functions of orders $0$ and $1$, which we denote here by $\psi(0,\cdot)$ and $\psi(1,\cdot)$ respectively. The function $\psi(1,\cdot)$ is strictly positive and decreasing on the entire positive real axis (see e.g.~\cite[Theorem 1.2.5]{AAR}), which implies the functions $|\psi(0,t)|$ and $|\Gamma(t)|$ are decreasing for $0 < t \leq t_0$ and increasing for $t > t_0$, where $t_0 \approx 1.46$ is the positive zero of $\psi(0,\cdot)$, which has to be taken into account when estimating the derivatives involving $\Gamma$. For $\ell_2 > 2$, we use \eqref{eq:univariate-mixed} and \eqref{eq:univariate-inf-mixed} only with $n = -1$ and $n = 0$ respectively, avoiding the need to to estimate these derivatives at all.
	
	Secondly, the integrals appearing in \eqref{eq:univariate-mixed-r} and \eqref{eq:univariate-inf-r-mixed} depend on $\alpha$ (the latter also on $r$) in less than immediately obvious ways. For \eqref{eq:univariate-mixed-r}, we simply note that
	\begin{equation} \label{eq:integrand-alpha}
		t^{\frac{2n+3+2\lceil\frac{\ell_1}{2}\rceil}{\alpha}} \leq \max\Bigl( t^{\frac{2n+3+2\lceil\frac{\ell_1}{2}\rceil}{\alpha_-}}, t^{\frac{2n+3+2\lceil\frac{\ell_1}{2}\rceil}{\alpha_+}}\Bigr) \quad \forall t > 0,
	\end{equation}
	if $\alpha_- \leq \alpha \leq \alpha_+$, so it suffices to numerically precompute
	\[
	\int_{0}^{\infty} |\log(t)|^{\ell_2} \max\Bigl( t^{\frac{2n+3+2\lceil\frac{\ell_1}{2}\rceil}{0.5 + (i-1)\Delta}},t^{\frac{2n+3+2\lceil\frac{\ell_1}{2}\rceil}{0.5 + i\Delta}} \Bigr) t^{-1}|p_{\ell_2}(t)| \rme^{-t} \rmd t,
	\]
	for $\ell_2 \in 1{:}4$, $\lceil\frac{\ell_1}{2}\rceil \in \{0,1\}$ and $i \in 1{:}i^*$, and use each of these for in the respective square $Q_{i,j}$. For numerical integration, we use the Julia library \texttt{QuadGK.jl}.
	
	Concerning integrals of the form \eqref{eq:univariate-inf-r-mixed}, we recall that parameter $\beta_1 = \pi + \alpha\varphi$ depends on $\alpha$, and note that it possible to write
	\begin{equation}\label{eq:M-derivatives}
		\frac{\partial^{\ell_2}}{\partial \alpha^{\ell_2}}\bigl[ \rme^{\rmi(n+1)\beta_1} \tau^{(n+1)\alpha}M_{n+1}\bigl(\rme^{\rmi \beta_1}\tau^{\alpha}\bigr)\bigr]
		= \tau^{(n+1)\alpha} \bigl(\log(\tau) + \rmi \varphi\bigr)^{\ell_2}\sum_{k=0}^{\ell_2} b_{n,k} \tau^{k\alpha} M^{(k)}\bigl( \rme^{\rmi \beta_1}\tau^{\alpha} \bigr),
	\end{equation}
	where the coefficients $b_{n,k}$ depend on $\alpha$ but their absolute values do not. For example, for $\ell_2 = 2$, the latter expression can be written as
	\begin{align*}
		\bigl( \rme^{\rmi \beta_1} \tau^\alpha \bigr)^{n+1} \bigl(\log(t) + \rmi \varphi\bigr)^{2}\Bigl( & (n+1)^2 M\bigl(\rme^{\rmi \beta_1}\tau^{\alpha}\bigr) - (2n + 3) \rme^{\rmi \alpha \varphi} \tau^{\alpha}M'\bigl(\rme^{\rmi \beta_1}\tau^{\alpha}\bigr) \\
		& \qquad + \bigl(\rme^{\rmi \alpha \varphi} \tau^{\alpha}\bigr)^2 M''\bigl(\rme^{\rmi \beta_1}\tau^{\alpha} \bigr)\Bigr).
	\end{align*}
	The crux of all this is that by applying the triangle inequality to \eqref{eq:M-derivatives} and using simple uniform estimates for $M$ and its derivatives (which is possible since $\rme^{\rmi \beta_1}\tau^{\alpha}$ by construction always has negative real part) , the integral appearing in \eqref{eq:univariate-inf-r-mixed} can be decomposed as a linear combination of integrals of the form
	\[
	\int_{0}^{\infty} \big|\log(\tau) + \rmi \varphi\big|^{\ell_2} \tau^{k\alpha + \ell_1} \rme^{\sin(\varphi) r\tau} \rmd t, \quad n+1 \leq k \leq n+\ell_2+1.
	\]
	Making the substitution $r \tau \eqdef t$, doing some elementary massaging for the resulting integrand, taking $\varphi \to -\pi_\alpha$ and using a discretization estimate similar to \eqref{eq:integrand-alpha}, we again end up with a finite collection of integrals that can be precomputed.
	
	The bivariate version of this estimate, \eqref{eq:bivariate-inf-mixed-r}, comes with the additional ingredient of the Hankel functions, which we use the Julia library \texttt{SpecialFunctions.jl} to compute.
	
	All in all, we have described several different ways to bound
	\begin{equation}\label{eq:square-sup}
		\sup_{(r,\alpha) \in Q_{i,j}} | f^{(\ell_1,\ell_2)}(r;\alpha) |,
	\end{equation}
	and for each square $Q_{i,j}$, we may take the smallest out of these bounds as the ultimate bound \eqref{eq:square-sup}.
	
	It then remains to estimate
	\[
	\inf_{(r,\alpha) \in Q_{i,j}} f(r;\alpha),
	\]
	from below. For this purpose, we have first precomputed $f(r;\alpha)$ at the corners of the $Q_{i,j}$'s using different numerical integration routines for different ranges of the parameters; see Section 3.1 of the main paper, where this is explained in the context of the spline approximation.
	
	We then observe that $f(r;\alpha)$ is for fixed $\alpha$ a decreasing function of $r > 0$, both in the univariate and in the bivariate case. Thus, writing $Q_{i,j} \eqdef [r_{-,j},r_{+,j}]\times[\alpha_{-,i},\alpha_{-,i}]$, we may use the fundamental theorem of calculus to obtain
	\begin{align*}
		\inf_{(r,\alpha) \in Q_{i,j}} f(r;\alpha) &= \inf_{\alpha \in [\alpha_{-,i},\alpha_{-,i}]} f(r_{+,j};\alpha) \\& \geq \min\bigl( f(r_{+,j};\alpha_{-,i}),f(r_{+,j};\alpha_{+,i}) \bigr) 
		- \frac{\Delta}{2} \sup_{(r,\alpha) \in Q_{i,j}} | f^{(0,1)}(r;\alpha) |,
	\end{align*}
	where the latter can further be estimated using the bounds discussed in the context of \eqref{eq:square-sup}. It turns out that with a small enough $\Delta$, such as $\Delta = \frac{2}{10^3}$, this yields a fair lower bound for a function like $f(r;\alpha)$ which is otherwise very difficult to bound from below.
	
	\section*{Tail error bounds}
	
	Here we establish relative error bounds for the series decompositions \eqref{eq:univariate-inf-expansion-1} and \eqref{eq:nolan-asymptotic}, used in the main paper with $n = 3$ and $r > 30$.
	
	First, in the univariate case, we can rewrite \eqref{eq:univariate-inf-expansion-1} as
	\begin{align*}
		f(r;\alpha) & = \sum_{k=1}^{3} \frac{(-1)^{k+1}\Gamma(k\alpha+1)\sin(\frac{k\alpha\pi}{2})}{\pi k!} r^{-k\alpha-1} + \mathcal{R}_4(r;\alpha) \\
		& \qquad \qquad \eqdef \sum_{k=1}^{3} c^{\mathcal{S}}_k(\alpha) r^{-k\alpha-1} + \mathcal{R}_4(r;\alpha) \eqdef \mathcal{S}_3(r;\alpha) + \mathcal{R}_4(r;\alpha),
	\end{align*}
	with
	\[
	|\mathcal{R}_4(r;\alpha)| \leq \frac{\Gamma(4\alpha+1)}{24\pi \sin(\pi_\alpha)^{4\alpha+1}}r^{-4\alpha-1} \defeq c^{\mathcal{R}}_4(\alpha) r^{-4\alpha-1}.
	\]
	Thus, for $r \geq 30$, we have the following estimates:
	\begin{align*}
		\Big|1 - \frac{f(r;\alpha)}{\mathcal{S}_3(r;\alpha)}\Big| = \Big|\frac{\mathcal{R}_4(r;\alpha)}{\mathcal{S}_3(r;\alpha)}\Big| & \leq \frac{c^{\mathcal{R}}_4(\alpha) r^{-4\alpha-1}}{|c^{\mathcal{S}}_1(\alpha)|r^{-\alpha-1} - |c^{\mathcal{S}}_2(\alpha)|r^{-2\alpha-1} - |c^{\mathcal{S}}_3(\alpha)|r^{-3\alpha-1}} \\
		& = \frac{c^{\mathcal{R}}_4(\alpha) r^{-3\alpha}}{|c^{\mathcal{S}}_1(\alpha)| - |c^{\mathcal{S}}_2(\alpha)|r^{-\alpha} - |c^{\mathcal{S}}_3(\alpha)|r^{-2\alpha}},
	\end{align*}
	assuming these calculations are valid in the sense of the latter denumerator being positive for $r = 30$ -- numerical considerations show that this is indeed the case. The rightmost quantity is then a decreasing function of $r \geq 30$, and we thus get a uniform error bound by investigating it with $r = 30$:
	\[
	\sup_{r > 30, \; 0.5 \leq \alpha \leq 1.9} \,
	\Big|1 - \frac{f(r;\alpha)}{\mathcal{S}_3(r;\alpha)}\Big| \leq 0.00096.
	\]
	
	In the bivariate case, we may proceed similarly, this time with \eqref{eq:nolan-asymptotic} for $n = 3$ and bounding the remainder term with \eqref{eq:bivariate-inf-r}. We can estimate the integral in \eqref{eq:bivariate-inf-r} using grid-based numerical upper bounds like in the previous section. We thus find
	\[
	\sup_{r > 30, \; 0.5 \leq \alpha \leq 1.9} \,
	\Big|1 - \frac{f(r;\alpha)}{\mathcal{S}_3(r;\alpha)}\Big| \leq 0.0016.
	\]
	
	We may then use the logarithm function's basic continuity properties near $1$ to infer
	\[
	\sup_{r > 30, \; 0.5 \leq \alpha \leq 1.9} \, |\log\,f(r;\alpha) - \log\, \mathcal{S}_3(r;\alpha)| \leq
	\begin{cases}
		0.00097  & \quad \textrm{(univariate case);} \\
		0.0017   & \quad \textrm{(bivariate case).}
	\end{cases}
	\]

	\printbibliography
	
	
	
	
	
	
	
	
	
	

\end{document}